\documentclass[12pt]{article}
\usepackage[utf8]{inputenc}
\usepackage[english]{babel}
\usepackage{amsfonts}
\usepackage{textcomp}
\usepackage{amsmath, amssymb, graphicx, upgreek}
\renewcommand{\theequation}{\arabic{section}.\arabic{equation}}
\usepackage{geometry}
\usepackage{enumerate}
\usepackage{hyperref}
\usepackage{braket}
\usepackage{color}
\usepackage{cite}
\usepackage{float}
\usepackage{indentfirst}
\usepackage{epsfig}
\usepackage{booktabs}
\usepackage{subfigure}
\usepackage{caption}
\usepackage{ulem}
\usepackage{array}
\usepackage{multirow}

\newcommand{\be}{\begin{equation}}
\newcommand{\ee}{\end{equation}}
\newcommand{\bea}{\begin{eqnarray}}
\newcommand{\eea}{\end{eqnarray}}

\newcommand{\lb}{\label}

\newcommand{\fnm}{\footnotemark}
\newcommand{\fnt}{\footnotetext}

\begin{document}

\begin{center}

  \large \bf
  Holographic RG flows in a 3d gauged supergravity at finite temperature
\end{center}

\vspace{15pt}

\begin{center}

 \normalsize\bf
       Anastasia Golubtsova\fnm[1]\fnt[1]{golubtsova@theor.jinr.ru}$^{, a,b}$,  Alexander Nikolaev\fnm[2]\fnt[2]{alexn99@gmail.com}$^{,b,c}$,  Mikhail Podoinitsyn\fnm[3]\fnt[3]{mpod@theor.jinr.ru}$^{, a}$
 \vspace{7pt}

 \it (a) \ \ Bogoliubov Laboratory of Theoretical Physics, JINR,\\
Joliot-Curie str. 6,  Dubna, 141980  Russia  \\

(b) \  Steklov Mathematical Institute, Russian Academy of Sciences\\
Gubkina str. 8, 119991 Moscow, Russia

(c)\, Physics Department, Lomonosov Moscow State University, 1-2 Leninskie Gory, Moscow 119991, Russia

 \end{center}
 
  \vspace{15pt}

\begin{abstract}
In this paper we consider finite-temperature holographic RG flows in $D=3$ $\mathcal{N}=(2,0)$ gauged truncated supergravity  coupled to a sigma model with a hyperbolic target space. In the context of the holographic duality, fixed points (CFTs) at finite temperature are described by AdS black holes. We come from the gravity EOM to a 3d autonomous dynamical system, which critical points can be related to fixed points of  dual field theories. Near-horizon black hole solutions correspond to infinite points of this system. We use Poincar\'e transformations to project the system on $\mathbf{R}^3$ into the 3d unit cylinder such that the infinite points are mapped onto the boundary of the cylinder. We explore numerically the space of solutions. We show that the exact RG flow at zero temperature is the separatrix for asymptotically AdS black hole solutions if the potential has one extremum, while for the potential with three extrema the separatrices are RG flows between AdS fixed points. We find near-horizon analytical solutions for asymptotically AdS black holes using the dynamical equations. We also present a method for constructing full analytical solutions.

\end{abstract}	

\tableofcontents

\newpage

\section{Introduction}

The notion of renormalization group \cite{Gell-Mann:1954yli,Bogolyubov:1956gh,Bogolyubov:1959bfo,Symanzik:1970rt,Callan:1972uj,WilsonKogut} gives a unified approach to 
the systems with many degrees of freedom.  Fixed points of the renormalization group equations are conformal field theories, which are connected by RG flows. RG flows can be induced either by an operator deformation of the original CFT, or by an operator VEV.

Holographic duality provides a map between $d$-dimensional quantum field theories at strongly coupling and  weakly coupled $(d+1)$-dimensional gravity \cite{Maldacena:1997re,Witten:1998qj,Gubser:1998bc}. This map is furnished by calculations of observables from both sides. This procedure yields a technical difficulty, namely, both field and gravity sides suffer from infinities, that can be observed, for example, from calculations of two-point correlation functions \cite{Arefeva:1998okd}. One needs to perform the renormalization procedure to remove divergences systematically and consistently.

In \cite{Akhmedov:1998vf} it was found that equations of motion of Type IIB supergravity on $AdS_{5}\times S^5$ can be associated with renormalization group flow equations of the dual gauge theory. 
In the holographic prescription RG flows \cite{deBoer:1999tgo, FGPW, deHaro:2000vlm} are described in terms of gravitational domain wall solutions with AdS boundaries and certain boundary conditions for the field content \cite{deHaro:2000vlm, deB,BFS}.  
 The extrema of the scalar potential are in a one-to-one correspondence with fixed points of RG flows, thus typically maxima of the potential are related with UV fixed points, and minima with IR fixed points.
While the asymptotic geometries of the domain wall solutions can be related with fixed points of RG flow (CFTs) of  dual field theories, the solutions themselves with certain boundary conditions  can be interpreted as  deformations of fixed points either by relevant operators or by non-zero vacuum expectation values of operators \cite{Bianchi:2001de,SkenderisLec,Papadimitriou:2004ap,Papadimitriou:2004rz,Papadimitriou:2007sj}.

In the extrema of the scalar potential the gravity solutions are AdS spacetimes, so as, it was said above, the holographic RG flows correspond to gravity solutions, which are asymptotically AdS. However, there can be an additional case, namely,  when the scalar potential has only one extreme point, this can be the case for supergravity actions after reduction from higher dimensions or truncation. Then the flow runs from the extremum of the potential to infinity as the scalar field diverges, $\phi\to\pm\infty$.  The corresponding geometry of the critical point is just scale-invariant and doesn't possess conformal invariance\footnote{In what follows we refer to them as scaling solutions.}.
The solution 
is singular since the divergence of the scalar field and the potential leads to the divergence of curvature invariants of the metric. However, such solutions can be acceptable in holography, if the Gubser's bound  \cite{Gubser:2000nd} and/or a criterion of spectral computability is satisfied.
A classification of holographic RG flows at zero temperature was given in \cite{Kiritsis:2016kog}, where exotic RG flows including those with bounces and skipping RG flows were also discussed.

To study thermal properties of QFT it is of interest to explore a generalization of holographic RG flows to finite temperature. The analysis of holographic RG flows at finite temperature allows to find out the consistence of the theory. From the gravity side, fixed points at finite temperature are described by AdS black holes. Thus, the domain wall solution is replaced by a black hole (brane) written in the domain wall coordinates.

In \cite{Gursoy:2018umf,Bea:2018whf} holographic thermal RG flows were studied to clarify properties of exotic RG flows using the holographic approach. Holographic RG flows with finite-temperature corresponding generalizations were investigated in works \cite{Gursoy:2008za,IYaRannu,AGP,IAHoReGR,IYaRa,Arefeva:2024vom} for applications to bottom-up models of holographic QCD.

Studies of holographic RG flows at finite temperature can clarify physics of a black hole interior.
In  \cite{Frenkel:2020ysx,Caceres:2022smh} it was considered  a deformation of the thermal state of a dual CFT by a relevant operator, which triggers a holographic RG flow at non-zero temperature. It was shown that  this flow can be continued smoothly through the horizon such that it deforms the Schwarzschild singularity into the more general Kasner universe. In \cite{Caceres:2023zft}  the perturbation of holographic RG flows at finite temperature in the context of \cite{Frenkel:2020ysx} by a shock wave was considered.

In this paper we consider fixed points of holographic RG flows at finite temperature for $D=3$ $\mathcal{N}=2$ gauged supergravity coupled to a sigma model with a hyperbolic target space $\mathbb{H}^2 = \frac{SU(1,1)}{U(1)}$ \cite{Sezgin, Deger:2002hv}. The supergravity model is truncated, such that we deal with a model of dilaton gravity, which scalar potential has either a single extremum defined at the origin or three extrema depending on the value of the target space curvature. Note that the vacuum at the origin of the potential is supersymmetric, while two additional are not. Holographic RG flows at zero temperature for this model were studied in \cite{Deger:2002hv,Golubtsova:2022hfk,NE-RG}. An exact half-supersymmetric holographic RG flow interpolating between AdS${}_{3}$ and scaling geometries was found in \cite{Deger:2002hv}. Despite this RG flow has a singularity as $\phi \to \infty$, the solution satisfies Gubser's bound for the case of the negatively defined scalar potential with a single extremum. In \cite{NE-RG} extending previous works \cite{Deger:2002hv,Golubtsova:2022hfk} holographic RG flows for the model were discussed choosing different boundary conditions for the field content.
In \cite{Park:2018ebm} the  behaviour of the c-function of the entanglement entropy along the RG flow for the holographic model \cite{Sezgin, Deger:2002hv} has been studied analytically and numerically.

Our primary interest will be generalization of the holographic RG flows at zero temperature from \cite{Golubtsova:2022hfk,NE-RG} to the finite temperature case. As in these works we come from the equations of motion for the holographic model to a 3d first-order autonomous dynamical system. This can be considered as an interpretation of RG flows of dual CFTs as the dynamical system\cite{Gukov}, such that RG flows are certain trajectories in the phase space. Zero-temperature holographic RG flows with Dirichlet boundary conditions have fixed points, which are equilibria of the dynamical system and are related with the maxima and minima of the scalar potential, correspond to  UV CFT and IR CFTs, respectively. These fixed points of the system have different types of the dynamical stability. For the scalar potential with a single extremum and Dirichlet boundary conditions RG flows have a UV conformal fixed point at the origin and IR scaling fixed points as the potential goes to infinity.  Infinite points of the dynamical system are related with black hole solutions near the horizon. 

We apply Poincar\'e transformations \cite{Lefschetz,Perko, Bautin} to project the phase space of the system into the unit cylinder, such that the infinite points are projected onto its boundary. Thus, near-horizon regions of asymptotically AdS black holes are mapped onto the critical line of the system  on the boundary of the cylinder.
Note, that near-horizon regions of black holes as infinite points of autonomous dynamical systems, which have origin from some gravity models were considered in works \cite{Ganguly:2014qia,Cruz:2017ecg}. The authors used Poincar\'e transformations to project the phase space of the gravitational dynamical system into the ball to explore infinite points of the equations of motion and, then, estimate asymptotic black hole  solutions. 

We solve numerically the equations to find black hole solutions starting from  near-horizon regions and ending at AdS zero-temperature fixed points. To analyze the numerical solutions, we also construct the known exact RG flow and AdS black hole solutions in the phase space, that helps us to understand the global phase portrait of the system.  The construction and analysis of the solutions are similar to  \cite{AGP,Arav:2020asu}. For the negatively defined scalar potential with one extremum the black hole solutions tending to the AdS fixed point pass very close near the curve of the exact RG flow. Thus, this is a separatrix for the asymptotically AdS black hole solutions. However, for the solutions corresponding to the potential with two additional minima this no longer holds. For this case separatrices of the asymptotically AdS black hole solutions are the flows, which connect the AdS fixed points at the maximum and minima of the potential and dual to different CFTs. 

It is interesting to note that from the point of view of the dynamical system holographic RG flows at zero temperature are trajectories between certain critical points, while in the thermal case RG flows are hypersurfaces, which one of the boundaries is a separatrix for asymptotically AdS black hole solutions with different $\phi_{h}$.


While to construct the full analytic black hole solutions with AdS asymptotics is quite difficult, we are able to find a description for the near-horizon region.  The solutions are characterized by the value of the scalar field at the horizon $\phi_{h}$ and can be divided into two classes: 1) when $\phi_{h}$ matches with a value of the critical point of the scalar potential and 2) when $\phi_{h}$ is located between of the critical points of the potential or between the critical point and the zero of the potential.

For flows starting near the maximum of the scalar potential, we are able to improve the analytical black hole solutions\footnote{In this work we restrict ourselves to the solution for the scalar field as a function of the scale factor.} in sense that the solutions can be extended to the boundary as well.

The paper is organized as follows. In section \ref{sec:setup} we give a setup: we review the holographic model and derive the 3d autonomous dynamical system from the equations of motion. In section \ref{sec:cylmap} using Poincar\'e transformation we project the system into the unit cylinder, that provides us to understand the global phase portrait and construct the flows numerically.  We analyze the behaviour of the flows from the near-horizon regions to the boundaries. In section \ref{sec:nhsol} we construct analytic near-horizon black hole solutions. In section \ref{sec:solPhiImp} we find a generalization of the solutions for the scalar field obtained in the previous section, which are valid for the region near the boundary. We finish with further  discussion in section \ref{sec:concl}.
Some technical details are left for Appendices.

\newpage

\setcounter{equation}{0}

\section{Setup}\label{sec:setup}
\subsection{The supergravity model}

The gravity model of our interest is $\mathcal{N}=(2,0)$  AdS$_3$ supergravity with $n$-copies of $\mathcal{N}=(2,0)$ scalar multiplet designed in \cite{Sezgin}.
The field content includes a vielbein $e_{\mu}^{\,a}$, a gravitini $\psi_{\mu}$ and a gauge field $A_{\mu}$. 
The scalar multiplet consists of a complex scalar field $\Phi$ and by a doublet of spinorial fields $\lambda$. In the work \cite{Sezgin} the generic case where scalars parametrize a coset space $G/H$ was discussed with a compact or non-compact $G$ and its maximal compact subgroup $H$. In this paper, as in \cite{Golubtsova:2022hfk,NE-RG}, we focus on the non-compact case for $G$ with a single complex scalar field $\Phi$, so the coset space is  $\frac{SU(1,1)}{ U(1)} =\mathbb{H}^2$.

The bosonic part of the Lagrangian \cite{Sezgin} reads 
\be \label{lag1}
    e^{-1}\mathcal{L} = \frac{1}{4} R - \frac{e^{-1}}{16 m \, a^4} \epsilon^{\mu\nu\rho}A_{\mu} \partial_{\nu} A_{\rho} - \frac{|D_{\mu} \Phi|^2}{a^2(1- |\Phi|^2)} -   V(\Phi), 
\ee
where we define $e = \det e_{\mu}^{\,a}$, $D_{\mu}\Phi= (\partial_{\mu}+iA_{\mu})\Phi$ and 
 $-4m^2$ is the AdS$_3$ cosmological constant.  The
parameter $a$  is related to the curvature of the scalar manifold and should be non-zero\footnote{The $a=0$ case corresponds to the flat sigma model.}. 
The potential $V(\Phi)$ in \eqref{lag1} is given by
\begin{equation}
    V(\Phi) =  2 {m^2} C^2 \left(2 a^2 |S|^2 -  C^2 \right) \, ,
\label{potPhiC}
\end{equation}
where
\begin{equation}
    C = \frac{1 + |\Phi|^2}{1 - |\Phi|^2}, \quad     S = \frac{2 \Phi}{1 - |\Phi|^2}.
\end{equation}

 Introducing the following redefinition of the scalar field
\begin{equation}
        C \equiv \cosh \phi \,\,, \,\, |S| \equiv \sinh \phi
\end{equation}
allows us to rewrite  the kinetic term of the scalar field in the standard form, i.e. the Lagrangian (\ref{lag1}) comes to
\be\label{fullmodel}
    e^{-1}\mathcal{L} = \frac{1}{4} R - \frac{e^{-1}}{a^4} \epsilon^{\mu\nu\rho}A_{\mu} \partial_{\nu} A_{\rho} - \frac{1}{4a^2} \partial_{\mu} \phi \partial^{\mu} \phi
    - \frac{1}{4a^2} |S|^2 (\partial_{\mu} \theta + A_{\mu})(\partial^{\mu} \theta + A^{\mu}) -  V(\phi).
\ee

Switching off $\theta$ and the vector field $A_{\mu}$, $A_{\mu} = \theta = 0$, which is a consistent truncation of the theory, one finally comes to the action
\be\label{act}
S = \frac{1}{16\pi G_{3}} \int d^{3}x\sqrt{|g|}\left(R -\frac{1}{a^2}(\partial\phi)^2 -V(\phi)\right) + \frac{1}{8\pi G_{3}}\int_{\partial M} d^2 x \sqrt{|\gamma|}K,
\ee
where the potential $V(\phi)$  is given by:
\be\label{pot}
V(\phi)=2\Lambda_{uv} \cosh^2\phi\left[(1-2a^2)\cosh^2\phi+2a^2\right],
\ee
where $\Lambda_{uv} = -4 m^2$.

 The GHY term in (\ref{act}) includes the determinant of the induced metric  $\gamma$ on the boundary $\partial M$ and the extrinsic curvature $K$
 \be
 K =\gamma^{\mu\nu}K_{\mu\nu}, \quad K_{\mu\nu} =-\nabla_{\mu}n_{\nu}=\frac{1}{2}n^{\rho}\partial_{\rho}\gamma_{\mu\nu}, \quad n^{\mu}=\frac{\delta^{\mu}_{w}}{\sqrt{g_{ww}}}.
 \ee
 
\begin{figure} [h!]
    \centering
	\includegraphics[width= 7cm]{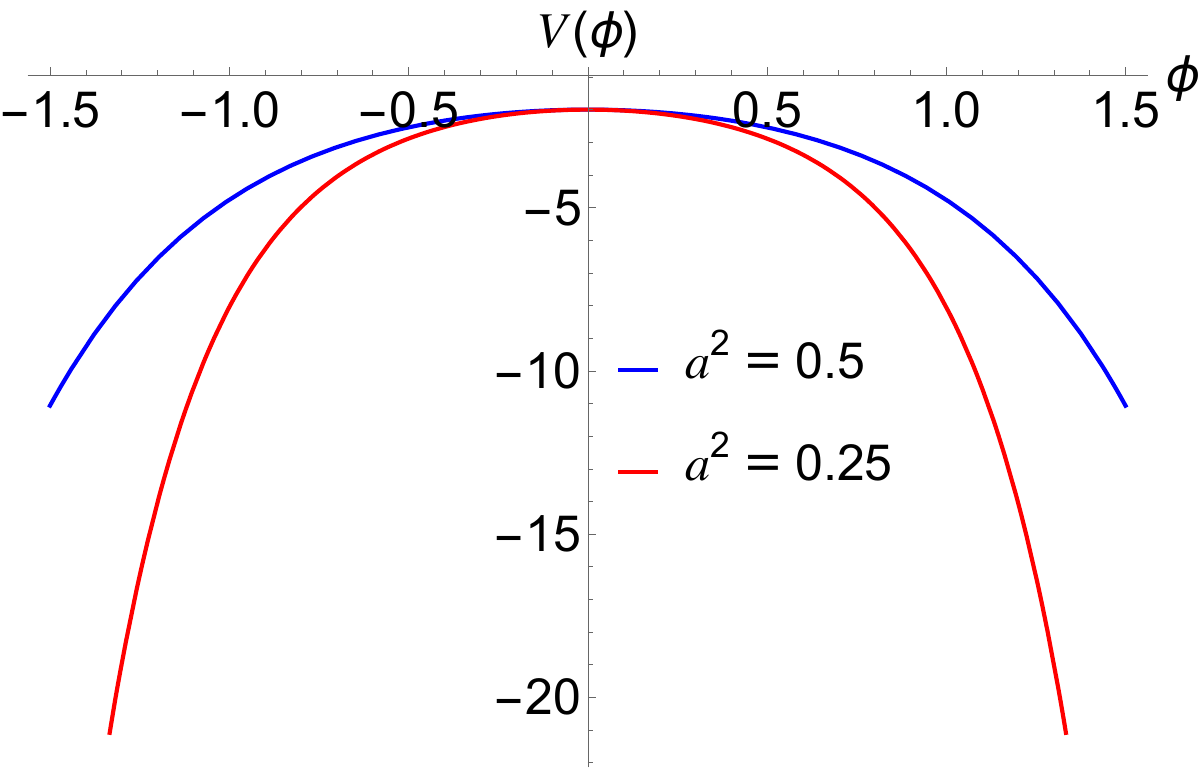} {\bf a)}\quad \includegraphics[width= 7cm]{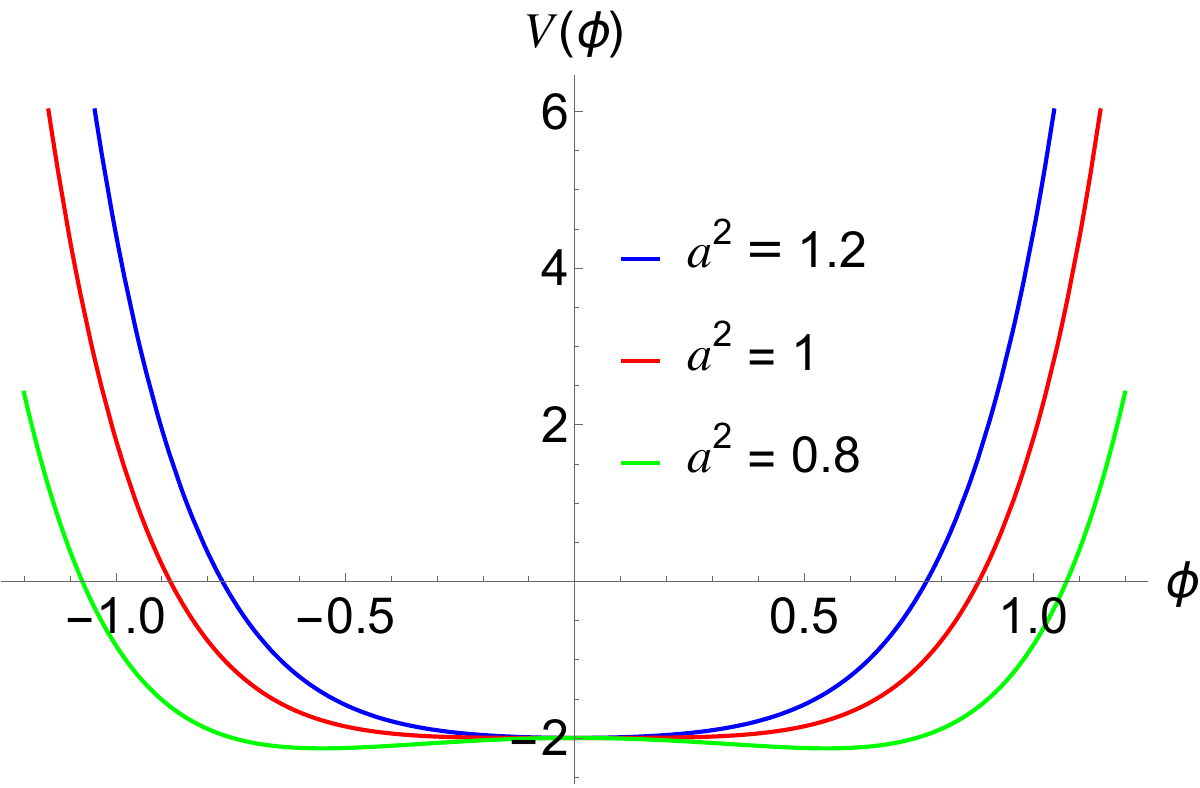} {\bf b)}
	\caption{\small The potential (\ref{pot}) against the scalar field $\phi$ for different values of $a$. For all we set $\Lambda_{uv}=-1$.  }
	\label{fig:scpotential}
\end{figure}

  We show the behaviour of the potential on the scalar field in Fig.~\ref{fig:scpotential}.  Note that for $a^2\leq \frac{1}{2}$ and for $a^2\geq 1$ the potential has one extremum,  namely, 
  \be\lb{phi10}
  \phi_{1} =0,
  \ee while for $\frac{1}{2}<a^2<1$ the  potential has two more extremal points, i.e.
 \be \lb{crP-v}
 \phi_{2,3} = \frac{1}{2}\ln\left(\frac{1\pm 2|a|\sqrt{1-a^2}}{2a^2 -1}\right).
 \ee
Moreover, if we consider the case $a^2>\frac{1}{2}$ the scalar potential has also zeroes:
\be\label{scpV0}
\phi_{\pm} = \pm \cosh ^{-1}\left(\frac{a}{\sqrt{a^2-\frac{1}{2}}}\right).
\ee
 We marked by $\phi_{3}$  and $\phi_2$ (\ref{crP-v}) the left and right extrema of the potential, correspondingly. In this paper we mainly discuss the cases 1) $0<a^2\leq\frac{1}{2}$ and 2) $\frac{1}{2}<a^2<1$.

The potential is related to the superpotential through the equation
\be\label{superPot}
V = \frac{a^2}{4}W'^2 -\frac{1}{2}W^2.
\ee
The exact supersymmetric solution for $W$ from eq. \eqref{superPot} has the following form
\be\label{supePsusy}
W_{\rm susy} = -\cosh^{2}\phi.
\ee
Note, that eq.(\ref{superPot}) has other solutions, which, however, cannot be obtained from  the supersymmetric variations of the model, comparing to (\ref{supePsusy}). Below we will call such solutions to (\ref{superPot}) as fake superpotentials.

The generic form  of the stress-energy tensor for (\ref{act}) reads as follows
\be\label{STE}
T_{\mu\nu}=\frac{1}{a^2}\left(\partial_{\mu}\phi\partial_{\nu}\phi-\frac{1}{2}g_{\mu\nu}\partial_{\sigma}\phi\partial^{\sigma}\phi\right)-\frac{1}{2}g_{\mu\nu}V.
\ee

To study a black hole solution for the model (\ref{act}) we consider the following ansatz of the  metric in the domain wall coordinates 
\be\label{metricmain}
ds^2 = e^{2A(w)}\left(-f(w)dt^2+dx^{2}\right) +\frac{dw^2}{f(w)},
\ee
where the radial coordinate runs the region $w\in (w_{h},+ \infty)$, with a location  of the horizon $w_{h}$ and the conformal boundary at  $w\to +\infty$. In (\ref{metricmain}) $A=A(w)$ is the scale factor, the function $f(w)$ is increasing on $(w_{h},\infty)$ and defined  the position of the horizon 
\be\label{horizonfunc}
f(w) \approx \begin{cases}
     1, & \text{if $w\to +\infty$}.\\
    0, & \text{if $w\to w_{h}$}.
  \end{cases}
\ee
The temperature of the solution (\ref{metricmain}) can be found as
\be\lb{HTemp}
T_{H} = \frac{e^{A(w_{h})}}{4\pi}\Bigl|\frac{df}{dw}\Bigr||_{w= w_{h}}.
\ee

The Hawking temperature (\ref{HTemp}) of the black hole  (\ref{metricmain}) corresponds to the temperature of the dual field theory.

For the scalar field  we also choose the dependence only on the radial  coordinate  $w$
\be
\phi = \phi(w).
\ee
We suppose that the scalar field takes some finite value at the horizon $w_{h}$
\be
\phi(w_{h}) = \phi_{h}.
\ee

Holographic RG flows and its  fixed points for the gravity model (\ref{act})-(\ref{pot}) with  an ansatz of the metric without a black hole, i.e. for (\ref{metricmain}) with $f=1$, was studied in \cite{Golubtsova:2022hfk,NE-RG}.

In what follows it is convenient to introduce the function
\be\label{funcgmain}
g = \ln f.
\ee

Then the Einstein equations to (\ref{act}) can be written down as follows
\bea
2\ddot{A}+2\dot{A}^2+\dot{A}\dot{g}+e^{-g}V+\frac{\dot{\phi}^2}{a^2}&=&0,\label{eom1}\\
3\dot{A}\dot{g}+2\ddot{A}+2\dot{A}^2+\dot{g}^2+\ddot{g}+e^{-g}V+\frac{\dot{\phi}^2}{a^2}&=&0,\label{eom2}\\
\dot{A}\dot{g}+2\dot{A}^2+e^{-g}V-\frac{\dot{\phi}^2}{a^2}&=&0,\label{eom3}
\eea
where we denote by dot the derivative with respect to the radial coordinate $w$. 
 The  scalar field equation can be represented as
\bea
\Box\phi= \frac{a^2}{2}V_{\phi}, \quad \Box=\frac{1}{\sqrt{|g|}}\partial_{\mu}\left(g^{\mu\nu}\sqrt{|g|}\partial_{\nu}\right)=e^{-2A}\partial_{w}\left(f(w)e^{2A}\partial_{w}\right),
\eea
or
\bea\label{eqd}
\dot{g}\dot{\phi}+2\dot{A}\dot{\phi}+\ddot{\phi}- \frac{a^2}{2}e^{-g}V_{\phi}=0,
\eea
where $V_{\phi}$ is a derivative of the scalar potential with respect to the scalar field.

For  values $\phi_h$ of the scalar field $\phi$ such that $V_{\phi}(\phi_{h})=0$ (i.e. $\phi_h =\phi_{1,2,3}$) the equations 
(\ref{eom1})-(\ref{eqd}) enjoy the exact AdS black hole solution (see  e.g. \cite{Gursoy:2018umf})
\be \lb{ck-sol}
A(w) = \sqrt{-\frac{V(\phi_{h})}{2}} \, w \,, \quad f(w) := e^{g(w)} = 1-e^{-\sqrt{-2 \, V(\phi_{h})} (w-w_h)},
\ee
where $V(\phi_{h})$ is the value of the potential at the extremum and the conformal boundary corresponds to the radial coordinate tending to infinity $w\to +\infty$.
Similar to pure AdS solutions we have only one AdS black hole at $\phi=\phi_{1}$ for $a^2\leq \frac{1}{2}$, while for $\frac{1}{2}<a^2< 1$ we have three AdS black hole solutions for different extrema of the scalar potential, i.e. at $\phi=\phi_{1,2,3}$. 

Note, that for $f=1$ the model has an exact half-supersymmetric solution, which interpolates between AdS and scaling geometries \cite{Deger:2002hv}. We briefly discuss  it in Appendix \ref{app:AppA}. 

\subsection{The scalar field near the boundary}

The scalar potential $V$ (\ref{pot}) near the extremum $\phi = \phi_{1}$ has the following expansion up to the quadratic order
\be\label{Vexp1}
V  = -2m^2 + 4m^2(a^2-1)\phi^2 + \mathcal{O}(\phi^3),
\ee
while near the other extrema $\phi =\phi_{2,3}$
\be\label{Vexp23}
V = -\frac{2a^4m^2}{2a^2-1}-\frac{8a^2(a^2-1)m^2}{2a^2-1}(\phi - \phi_{2,3})^2 +\mathcal{O}(\phi^3).
\ee
The  geometries corresponding to the extrema of the potential  $\phi_{1,2,3}$ are AdS spacetimes with different radii $\ell=\sqrt{-\frac{2}{V(\phi_{i})}}$ with $i=1,2,3$. So we assume that the  field theories dual to them are different 2d CFTs, which we call in what follows CFT$_{1}$, CFT$_{2}$, CFT$_{3}$, correspondingly. They are  characterized by different central charges, that can be seen from the Brown-Henneaux formula $c=\frac{3\ell}{2G}$.

In the regions of the extrema the solution to the equation of motion for the scalar field with (\ref{Vexp1}) (or (\ref{Vexp23}) has the asymptotic form
\be\label{AsympPhiB}
\phi  = \phi_{-} e^{-\Delta_{-}w} +\phi_{+}e^{-\Delta_{+}w},
\ee
with $w \to +\infty$.
In (\ref{AsympPhiB})  $\phi_{-}$ and $\phi_{+}$ are related to the source and to the VEV of the dual operator $\braket{\mathcal{O}}$, correspondingly.

From the scalar field equation with (\ref{Vexp1})-(\ref{Vexp23}) the conformal dimensions $\Delta$ of the operators of the dual field theories are  given by
\bea\label{confDim}
&&\phi_{1}:\qquad \Delta_{\pm} =1 \pm |1-2a^2|,\\
&&\phi_{2,3}: \qquad  \Delta_{\pm} =1 \pm \sqrt{1+\frac{8a^4(1-a^2)}{2a^2-1}}.
\eea
The dual field theory is two-dimensional, so
the operators are relevant for $\Delta < 2$, marginal for $\Delta = 2$ and irrelevant for $\Delta > 2$. The unitarity bound reads as $\Delta\geq 0$.

 For $\phi_{1}$ with
 $0<a^2 <1$:  we have $1\leq\Delta_{+}<2$ and $0<\Delta_{-}<1$, so there is an RG flow related with a deformation by a relevant operator both $\Delta_{+}$ and $\Delta_{-}$ are driven by relevant operators.
 
As for $\phi_{2,3}$ the conformal dimensions $\Delta_{\pm}$  with $\frac{1}{2}<a^2<1$ are such that the dual operator is irrelevant $\Delta_{+}>2$ and $\Delta_{-}<0$, so it breaks the unitarity bound.

The superpotential (\ref{superPot}) can be
expanded in series near $\phi=0$ 
\be \label{Wexp}
W_{\pm}(\phi)  = -\left(1+\frac{1}{2a^2}\Delta_{\pm}(\phi - \phi_*)^2 + \ldots \right),
\ee
where the choice of the branch $W_{+}$ of $W_{-}$ depends on whether the scalar field behaves near the conformal boundary as $e^{-\Delta_{+}r}$ or $e^{-\Delta_{-}r}$ (\ref{AsympPhiB}).
The fake superpotential can be also represented as an expansion in series near some critical point $\phi=\phi_{*}$ in the following form
\be\label{Wexp2}
W_{\pm}(\phi)  = -\sqrt{-2V_*} - \frac{1}{2a^2}\Delta_{\pm}(\phi - \phi_*)^2 + \ldots ,
\ee
with the conformal dimensions given by (\ref{confDim}).
Particularly, near $\phi=0$ the fake superpotential has the form \cite{NE-RG}
\bea
a^{2}<\frac{1}{2}\quad
&\begin{cases}
W_{+}= -1-\frac{1-a^2}{a^2}\phi^2+\mathcal{O}(\phi^3),\\
W_{-}=-1-\phi^2+\mathcal{O}(\phi^3),
\end{cases}\\
a^{2}>\frac{1}{2}\quad
&\begin{cases}
W_{+}=-1-\phi^2+\mathcal{O}(\phi^3),\\
W_{-}=-1-\frac{1-a^2}{a^2}\phi^2+\mathcal{O}(\phi^3),
\end{cases}\\
a^{2}=\frac{1}{2},\quad
&W_{\pm}=-1-\phi^2+\mathcal{O}(\phi^3).
\eea

In Appendix~\ref{app:AppC2} we give a list of possible holographic RG flows at zero temperature with different boundary conditions imposed. 

\subsection{Autonomous dynamical system}
In this paper it is of our interest to explore holographic RG flows of the model (\ref{act}) at finite temperature by means of the dynamical system theory. For this we
represent  equations of motion (\ref{eom1})-(\ref{eom3}), (\ref{eqd}) as an autonomous dynamical system. Then,  we explore the space of the solutions to the system numerically and construct  analytic black hole solutions near horizon.
To come to the dynamical system, we introduce new variables \cite{Gursoy:2008za,AGP,IAHoReGR}
\be\label{XvarDef}
X=\frac{d\phi}{d A} = \frac{\dot{\phi}}{\dot{A}},\qquad Y=\frac{d g}{d A}=\frac{\dot{g}}{\dot{A}},
\ee
where $g=\ln f$,  $\phi$ is the dilaton,  $A$ is the scale factor, and the derivatives are taken with respect to the radial coordinate $w$.

To parametrize the scalar field it is convenient to use the following variable:
\be\label{Zvar}
z = \frac{1}{1+e^{\phi}},
\ee
so that $z\in[0,1]$ for $\phi \in(-\infty;\infty)$. 

Doing some algebra with eqs.(\ref{eom1})-(\ref{eom3}),(\ref{eqd}) we find the following useful relations
\bea
\frac{\ddot{g}}{\dot{A}^2}&=&-Y^2-2Y,\\
\frac{\ddot{A}}{\dot{A}^2}&=&-\frac{1}{a^2}X^2,\\
\frac{\ddot{\phi}}{\dot{A}^2}&=&-2X-XY- \frac{a^2}{2}\frac{V_{\phi}}{V}\left(2+Y-\frac{X^2}{a^2}\right). 
\eea

The potential and its derivative are rewritten in terms of the variable $z$ (\ref{Zvar}):
\bea\label{V}
V&=&\Lambda \frac{(2 (z-1) z+1)^2 \left((2 (z-1) z+1)^2-2 a^2 (1-2 z)^2\right)}{8 (z-1)^4 z^4},\\ \label{Vpri}
V_{\phi}& = &\Lambda \frac{ \left(\left(1-2 a^2\right) \left((z-1)^8-z^8\right)-2 z^6 (z-1)^2+2 z^2
   (z-1)^6\right)}{2 (z-1)^4 z^4}.
\eea
For further calculations it is convenient to define the following function
\be \lb{CZa-in}
\mathcal{C}_{(z,a)} := \frac{a^2}{2} \frac{V_{\phi}}{V},
\ee
where with (\ref{V})-(\ref{Vpri}) we have
\be\label{VphV}
\frac{V_{\phi}}{V}= \frac{4 \left(\left(1-2 a^2\right) \left((z-1)^8-z^8\right)-2 z^6 (z-1)^2+2 z^2
   (z-1)^6\right)}{(2 (z-1) z+1)^2 \left((2 (z-1) z+1)^2-2 a^2 (1-2 z)^2\right)}.
\ee

The function $\mathcal{C}_{(z,a)}$  (\ref{CZa-in}) has the following  zeroes 
\be \lb{ext-in-z}
z_{1}= \frac{1}{2}, \quad z_{2,3}=\frac{1}{2} \left(1\mp \sqrt{\frac{1-3 a^2+2 \sqrt{a^2 \left(2
   a^2-1\right)}}{a^2-1}}\right),
\ee
which are obviously the extrema of the potential (\ref{phi10}) (\ref{crP-v}) in terms of the new  variable $z$.
The function (\ref{CZa-in}) is divergent for $a^2>1/2$ at the points 
\be \lb{zer-in-z}
z_{\pm}=\frac{1}{2} \left(1\pm\sqrt{4 a^2-2a \sqrt{4 a^2-2}-1}\right) \,, \quad 
\ee
which correspond to the zeroes of the scalar potential (\ref{scpV0}).

Then the equations of motion (\ref{eom1})-(\ref{eom3}),(\ref{eqd}) for the model (\ref{act}) are brought to the following ones
\bea\label{dz/da}
&&\frac{dz}{dA}=z(z-1)X,\\ \label{dx/da}
&&\frac{dX}{dA}=\left(\frac{X^2}{a^2}-Y-2\right)\left(X+\mathcal{C}_{(z,a)} \right),\\ \label{dy/da}
&&\frac{dY}{dA}=Y\left(\frac{X^2}{a^2}-Y-2\right).
\eea

 The equilibrium points are defined by zeros of the LHS of the system (\ref{dz/da})-(\ref{dy/da}).
We  distinguish three  cases  {\bf a)} $Y = 0$, {\bf b)} $Y\neq 0$ and {\bf c) $Y\to \infty$}. The critical points {\bf{a)}} with $Y=0$ correspond to zero-temperature regime. The stability analysis of them  was done in recent works \cite{Golubtsova:2022hfk,NE-RG}. The corresponding geometries include both AdS and scaling spacetimes. However, only for $a^2\leq \frac{1}{2}$ the solutions with scaling obey  Gubser's bound. 
It will be shown below that the neighborhoods of certain critical points {\bf c)} with $Y \to \infty$ correspond to near-horizon black hole solutions with AdS asymptotics.


 The exact AdS black hole geometry (\ref{ck-sol}) can be found from eqs.(\ref{dz/da})-(\ref{dy/da}), that is shown in  Appendix~\ref{app:AppB}. To construct and analyze asymptotically AdS black holes it is more convenient to compactify the phase space.
 Below we apply Poincar\'e transformations to project
  the 3d dynamical system (\ref{dz/da})-(\ref{dy/da}) defined on $\mathbf{R}^3$ into the unit cylinder.
  This helps us to explore infinite points corresponding to near-horizon black hole solutions and explore the global phase portrait of the system.
 

 \setcounter{equation}{0}
\section{Global structure of solutions}\label{sec:cylmap}

In this section, we discuss the global structure of solutions for the 3d dynamical system (\ref{dz/da})-(\ref{dy/da}).
To understand the behaviour of flows at infinity we apply the so-called Poincar\'e projection \cite{Lefschetz,Perko,Bautin}, which maps the system into the unit cylinder.
This method  was used for the complete description of  global phase portraits of  fairly general gravitational models in works \cite{Skugoreva:2014ena, Ganguly:2014qia, Cruz:2017ecg}. Typically, Poincar\'e projection deals with all variables of a dynamical system mapping it onto a hemisphere (a projective plane). 
 However, it will be more convenient for us to use the incomplete Poincar\'e projection (only for the variables $X$ and $Y$), since the coordinate $z\in[0,1]$ is already compact.
 
 \subsection{Dynamics on a 2d Poincar\'e disk}\lb{pdis}

Before giving a full analysis of the 3d autonomous system, we focus on the 2d system, which can be obtained as a slice of the system (\ref{dz/da})-(\ref{dy/da}) for some fixed $z$.
So the dynamics of the system is reduced to a discussion of eqs. (\ref{dx/da})-(\ref{dy/da}) on  the  $XY$\!-plane,
since the function $\mathcal{C}_{(z, a)}$ given by (\ref{CZa-in}) with (\ref{VphV}) tends to be constant.

 The equilibrium sets of the reduced 2d dynamical system include 
 \begin{enumerate}
 \item the parabola defined by  the equation
\be\label{redparabola}
\frac{X^2}{a^2} - Y - 2=0,
\ee 
\item  the point $P_0$: $X=-\mathcal{C}_{(z,a)}, Y=0$ coming from
\be\label{maindot}
X + \mathcal{C}_{(z, a)} =0.
\ee
\end{enumerate}
The latter  corresponds to an AdS vacuum if $\mathcal{C}_{(z,a)}=0$.
\begin{figure}[h!]
    \centering
	\includegraphics[width= 8cm]{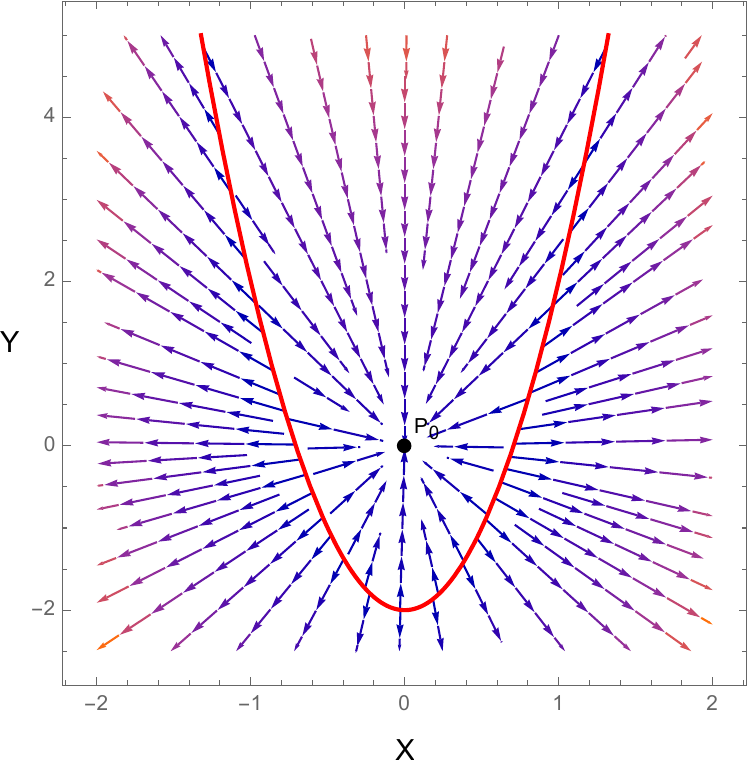}
	\caption{\small The phase portrait of the dynamical system (\ref{dx/da})-(\ref{dy/da}) at fixed $z$.  The critical curve is shown by red (\ref{redparabola}), while the critical point (\ref{maindot}) is shown by black. The parameters $a^2$ and $z$ are fixed as: $a^2 =0.25$ and $z=1/2$. }
	\label{Fig: VecFieInSys2d-p}
\end{figure}
The phase portrait of eqs.(\ref{dx/da})-(\ref{dy/da}) with fixed $z$ on the $XY$\!-plane is presented in Fig. \ref{Fig: VecFieInSys2d-p}, where the critical curve (\ref{redparabola}) is shown by red.

Note, the critical curve doesn't depend on $z$, in sense that this curve exists for each slice of $z=const$, while the location of the isolated critical point $P_0$ (\ref{maindot}) is defined by the value of $\mathcal{C}_{(z,a)}$.
The behavior of the trajectories inside the parabola in Fig.~\ref{Fig: VecFieInSys2d-p} is related with the position of the isolated critical point $P_0$.

To extract the information about the finite temperature flows we map  the system (\ref{dx/da})-(\ref{dy/da}) defined on the plane $\mathbf{R}^2$  into the disc $\mathbf{D}^2$. 
This allows to consider infinite points of eqs.(\ref{dx/da})-(\ref{dy/da}), which are difficult to be caught considering the system on $\mathbf{R}^2$.

The transformations include two parts, first, we map the dynamical system on the plane $\mathbf{R}^2$ onto the 2d unit sphere $\mathbb{S}^2$ and, then, we perform the perpendicular projection of the system on $\mathbb{S}^2$ into the 2d disc $\mathbf{D}^2$.
So we are led to study the dynamics of the system in the 2d disc $\mathbf{D}^2$.
 \begin{figure}[t]
    \centering
     \includegraphics[width=7.6cm]{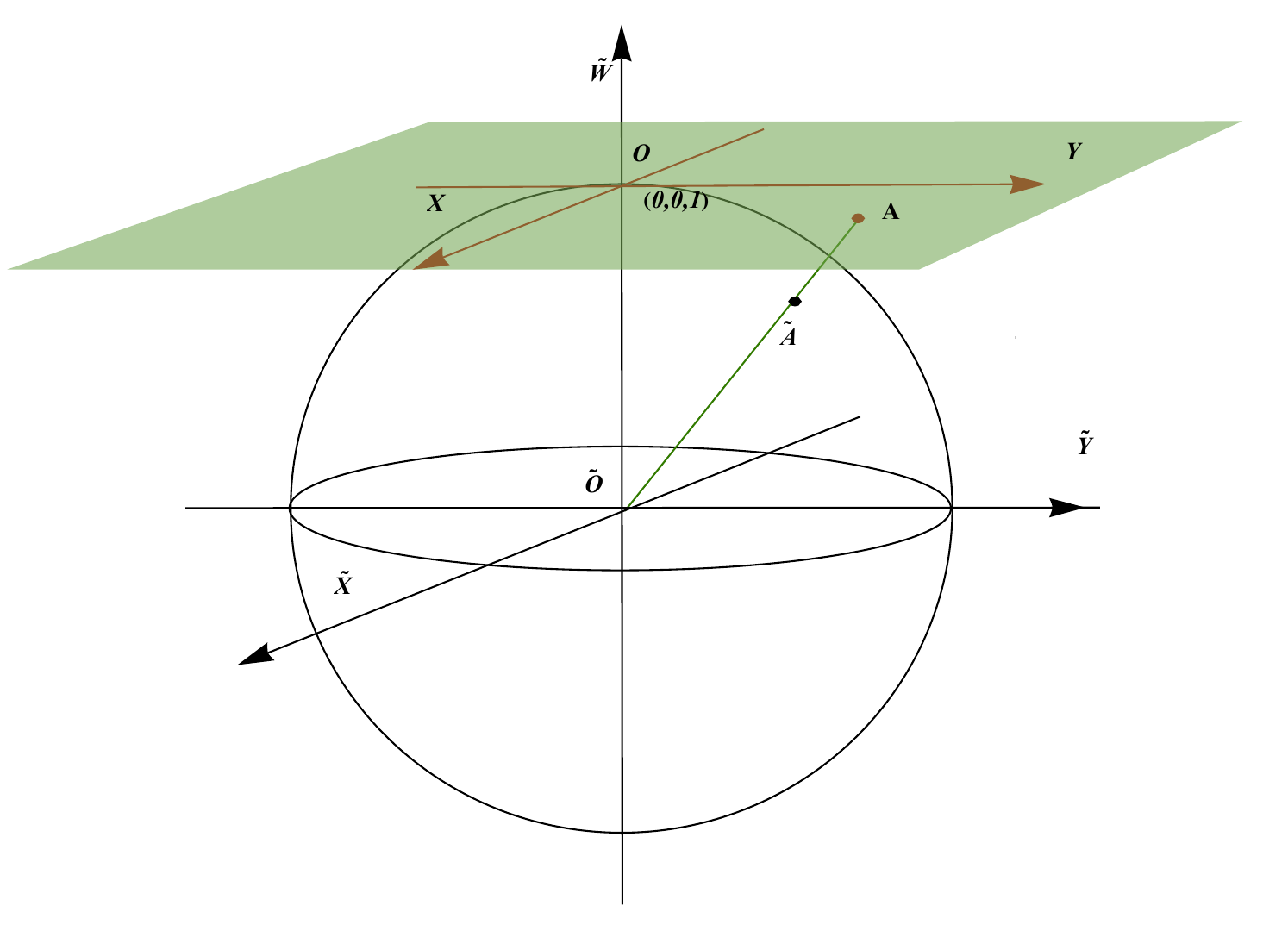}{\bf a)}
     \includegraphics[width=7.6cm]{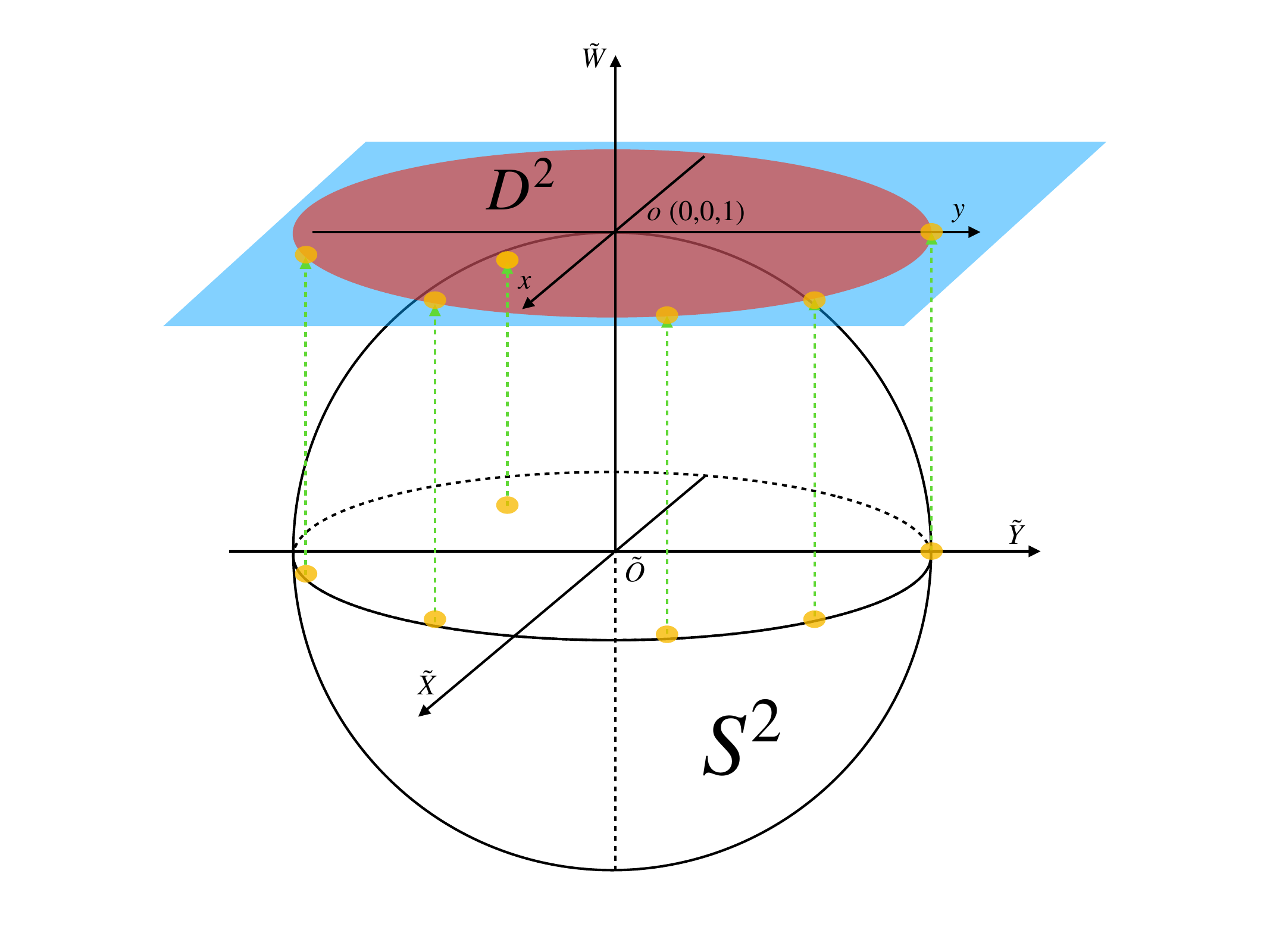}{\bf b)}
     \caption{\small {\bf a)}  The Poincar\'e map of the plane $\mathbf{R}^{2}$  ($OXY$)  onto the sphere $\mathbb{S}^2$ with coordinates ($\tilde{X}, \tilde{Y}, \tilde{W}$), where $O$ and $\tilde{O}$ - are the origins of the coordinates on the plane and sphere, correspondingly.  The point $O$ coincides with the north pole of the sphere $\mathbb{S}^2$, i.e. its coordinates are $(0,0,1)$. $\tilde{A} \in S^2$ is a projection of the point $A \in OXY$. {\bf b)} The transverse map of the system from the upper hemisphere of the unit sphere $\mathbb{S}^2$ into the unit disk $\mathbf{D}^2$. $\tilde{O}$ is the origin of the coordinates on $\mathbb{S}^2$, $o$ is the origin of the coordinates $(x,y)$ on  $\mathbf{D}^2$, that touches $\mathbb{S}^2$ at the point $(0,0,1)$.
     Yellow dots are images and preimages of the perpendicular projection.
     }
    \label{Fig: Poincare}
\end{figure}
To map the system on $\mathbf{R}^2$ (with coordinates $X$, $Y$) onto  $\mathbb{S}^2 \hookrightarrow \mathbf{R}^3$ (with coordinates $\tilde{X}$, $\tilde{Y}$, $\tilde{W}$) we use the Poincar\'e transformation ("the stereographic-like projection")  given by 
\begin{equation} \label{PoincareMap}
X = \frac{\tilde{X}}{\tilde{W}}, \quad Y = \frac{\tilde{Y}}{\tilde{W}},
\end{equation}
with
\be\label{sphereEmb}
\tilde{X}^2 + \tilde{Y}^2 + \tilde{W}^2= 1.
\ee
Then one can easily write down the coordinates of a new point on the sphere:
\be\label{revPoincareMap}
    \tilde{X} = \frac{X}{\sqrt{X^2 + Y^2  + 1}}, \quad
       \tilde{Y} = \frac{Y}{\sqrt{X^2 + Y^2 + 1}}, \quad     \tilde{W} = \frac{1}{\sqrt{X^2 + Y^2  + 1}}. 
\ee
We suppose that the plane $\mathbf{R}^2$, where the system (\ref{dx/da})-(\ref{dy/da}) is defined, touches the  sphere $\mathbb{S}^2$ at the north pole $(0,0,1)$.
 The points of  $\mathbf{R}^2$ are projected onto  $\mathbb{S}^2$ so that the points at infinity of the plane $\mathbf{R}^2$ come to the points on the equator of the sphere $\mathbb{S}^2$, as it is shown in Fig. \ref{Fig: Poincare} {\bf a)}. 
The relations (\ref{revPoincareMap}) define a one-to-one correspondence between the coordinates  on  the plane $X, Y \in \mathbf{R}^2$ and  the coordinates ($\tilde{X}, \tilde{Y}, \tilde{W}$) on the upper hemisphere of $\mathbb{S}^2$, i.e. $\tilde{W} > 0$.


Then we map the system on the sphere $\mathbb{S}^2$ into the transverse unit disk $\mathbf{D}^2$, which touches the sphere at $(0,0,1)$. Thus, the center of the unit disk coincides  with the point, at which it contacts  the plane, see Fig. \ref{Fig: Poincare} {\bf b)}.
The coordinate transformations for this map are given by the following formulae:
\begin{equation} \label{TransverseMap001}
        \tilde{X} = x, \quad \tilde{Y} = y, \quad \tilde{W} = \sqrt{1 - x^2 - y^2}. 
\end{equation}
Consequently, the system on the plane $\mathbf{R}^2$ is projected  into the disk by the coordinate transformations
\be \lb{ch-cor}
 X=\frac{x}{\sqrt{1-x^2-y^2}} \,, \quad Y=\frac{y}{\sqrt{1-x^2-y^2}}.
\ee
For the new coordinates $x, y$ on  $\mathbf{D}^2$ we also have the constraint
\be \lb{cyl-1}
x^2+y^2 \leq 1.
\ee
Eqs.(\ref{dx/da})-(\ref{dy/da}) take the form
\bea 
 \label{EqsonCyl0012n}
       x' &= &\mathrm{p}(x,y),  \\  
\label{EqsonCyl0013n}
           y'&= &\mathrm{q}(x,y), 
\eea
where the functions $\mathrm{p}$ and $\mathrm{q}$ are given by
\bea \nonumber 
\mathrm{p} &=&  \left( \sqrt{1-x^2-y^2} \left(2 \sqrt{1-x^2-y^2}+y\right)-\frac{x^2}{a^2} \right) \left(\mathcal{C}_{(z,a)}(x^2-1) -x \sqrt{1-x^2-y^2}\right) \,, \\ 
\label{q-disk2}
\mathrm{q} &=&  \left( \sqrt{1-x^2-y^2} \left(2 \sqrt{1-x^2-y^2}+y\right)-\frac{x^2}{a^2} \right)  \left(\mathcal{C}_{(z,a)} x -\sqrt{1-x^2-y^2}\right) y
\eea 
and we redefine the derivative as follows
\be \lb{derred}
\chi' = \sqrt{1-x^2-y^2}\, d\chi/dA.
\ee
The redefinition of the derivative (\ref{derred}) (see \cite{Lefschetz},\cite{Perko}) is admissible, since, it does not change the behavior of  flows, and allows us to avoid divergences, which arise for $x^2+y^2=1$.

Note that the line $y=0$ is an invariant manifold, so the flow doesn't leave it.

The critical sets of the system 
in the disk $\mathbf{D}^2$ (\ref{q-disk2}) are defined by zeroes of the RHS of eqs.  (\ref{EqsonCyl0012n})-(\ref{EqsonCyl0013n}), namely

\begin{enumerate}
\item the closed curve $e(a)$
\be\label{clcurv}
\sqrt{1-x^2-y^2} \left(2 \sqrt{1-x^2-y^2}+y\right)-\frac{x^2}{a^2} = 0,
\ee
\item the isolated point $p_0(z,a)$: 
\be\label{pointDisk}
x=-\frac{\mathcal{C}_{(z,a)}}{\sqrt{\mathcal{C}_{(z,a)}^2+1}},\quad y=0.
\ee
\end{enumerate}
Actually, there are two more points with coordinates $x=\pm 1$, $y=0$ at which the RHS of eqs.(\ref{EqsonCyl0012n})-(\ref{q-disk2}) vanish. These points appear as a result of the transformation (\ref{ch-cor}) and redefinition (\ref{derred}). They need to be taken into account to analyze the full 3d dynamical system, which we consider below.

The critical sets (\ref{clcurv})-(\ref{pointDisk}) are merely the initial critical sets (\ref{redparabola})-(\ref{maindot}) on $\mathbf{R}^2$ mapped into the disk $\mathbf{D}^2$. So the closed curve $e(a)$ (\ref{clcurv}) comes from eq.(\ref{redparabola}) by gluing the infinite ends of the parabola as a result of  Poincar\'e transformations \eqref{ch-cor}.%

\begin{figure} [t]
    \centering
	\includegraphics[width= 8cm]{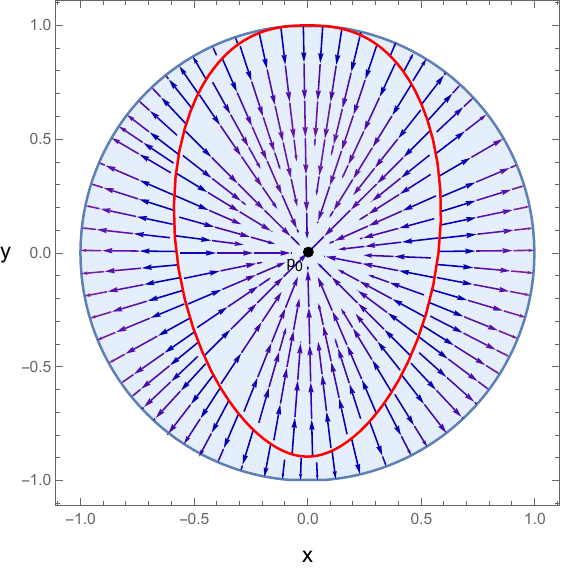}
	\caption{\small The flows of the system (\ref{EqsonCyl0012n})-(\ref{q-disk2}) with fixed $z$ in the disk $\mathbf{D}^2$.  The critical curve is shown by red, while the critical point $p_{0}$ is marked by black. The parameters $a^2$ and $z$  are set as: $a^2 =0.25$ and  $z=0.5$ (so $\mathcal{C}_{(z,a)}=0$). }
	\label{Fig: VecFieInSys2d}
\end{figure}

Classification of stability for the critical point $p_0$ at various $a$ and $z$ is given in  Table \ref{table:p02D-t}.  From this table we see that if $a^2\leq\frac{1}{2}$ the point $p_{0}$ is a stable node for all $z$, while for $a^2>\frac{1}{2}$ the type of stability for $p_{0}$ depends on the value of $z$ with respect to the intersections of the critical curve $e(a)$ and $p_{0}$. So  if  $z$ lies in the region between $z=0$ and the value of $z$ at the first intersection of $p_{0}$ and $e(a)$ ($z_{a}$) this is an unstable node, the same is valid for $z$ lying in the region between the value of $z$ at the second intersection of $p_{0}$ and $e(a)$ and $z=1$ ($z_{b}$), while if $z$ belongs to $z\in (z_{a},z_{b})$, $p_{0}$ is a stable node.

\begin{table}
\centering
\begin{tabular}  {|c|c|c|c|}
\hline & $0 \leq z \leq z_{a}$ & $z_{a}<z<z_{b}$ &  $z_{b} \leq z \leq 1$ \\
\hline $0 < a^2 \leq 1/2$ & none & stable node & none \\
\hline $ a^2> 1/2$ & unstable node & stable node & unstable node \\
\hline
\end{tabular}
\caption{The stability of the critical point $p_{0}$ for various values of $a^2$ and $z$; $z_{a}$ and $z_{b}$ are values of $z$, at which the point $p_0$ intersects the critical curve $e(a)$. 
}\lb{table:p02D-t}
\end{table}

\begin{figure}[h!]
    \centering
	\includegraphics[width= 7cm]{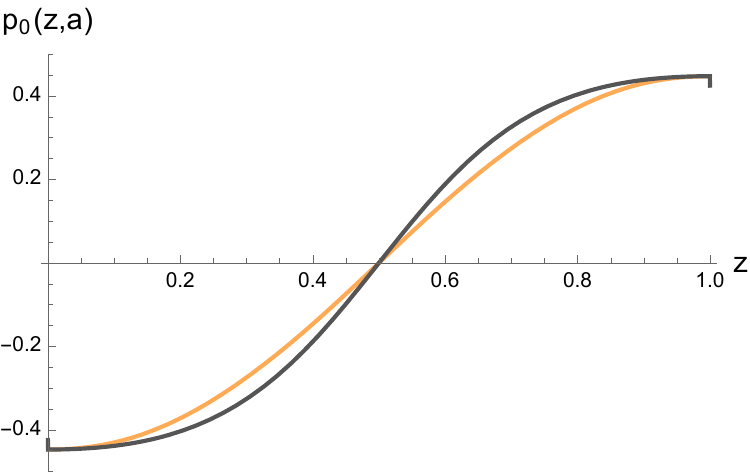}{\bf a)} \qquad \includegraphics[width= 7cm]{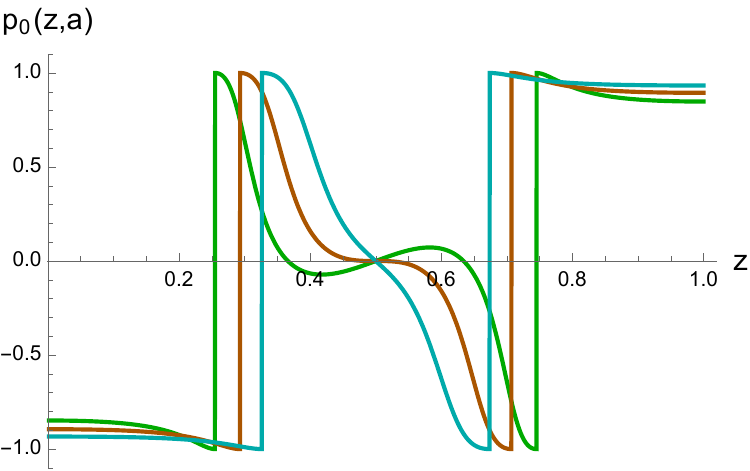}{\bf b)}\\
	\caption{\small The dependence  of the $x$-coordinate of $p_{0}$ on $z$ for various $a$: {\bf a)}$a^2=0.25$ by orange, $a^2=0.5$ by gray; {\bf b)} from left to right $a^2=0.8$, $a^2=1$, $a^2=1.2$, correspondingly.}
	\label{Fig: CrPCzPos}
\end{figure}

It's instructive to look at the behaviour of the $x$-coordinate of  $p_{0}$ on $z$. In Fig.~\ref{Fig: CrPCzPos} we depict the $x$-coordinate of $p_{0}$ on $z$ for different $a$. From Fig.\ref{Fig: CrPCzPos} {\bf a)} it can be seen that for $a^2=\frac{1}{2}$ the $x$-coordinate as the function of $z$ has a discontinuity in the boundary points $z=0$ and $z=1$. As for $a^2>\frac{1}{2}$ we observe from Fig.\ref{Fig: CrPCzPos} {\bf b)} that the $x$-coordinate of $p_{0}$ with $a^2=0.8$ vanishes for three values of $z$, which are related with the extrema of the scalar potential $V'_{\phi}=0$, it also has two extrema, which are related with $V''_{\phi}=0$.  For all $a^2>\frac{1}{2}$ the $x$-coordinate of $p_{0}$ has also a discontinuity at values of $z$, for which the scalar potential vanishes $V=0$.

For $a^2\leq \frac{1}{2}$ the critical point $p_{0}$ is always located inside of the critical curve. However, one should be careful with $a^2=\frac{1}{2}$ near $z=0$ and $z=1$, where the point $p_{0}$ jumps due to discontinuities, see Fig.~\ref{Fig: CrPCzPos} {\bf a)}. For $a^2=0.8$  $p_{0}$ can be both inside and outside of the critical curve. In this case, the point $p_{0}$ can also jump, that happens near regions for which $V=0$.

\begin{figure}[t]
  \centering
\includegraphics[width= 7.2cm]{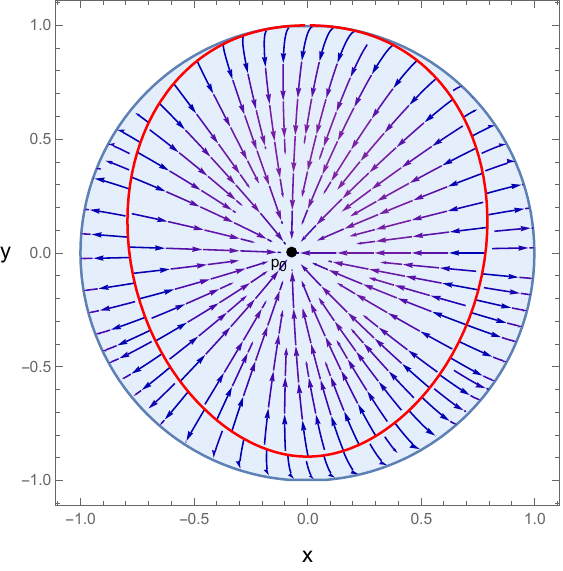}{\bf a)}\qquad
\includegraphics[width= 7.2cm]{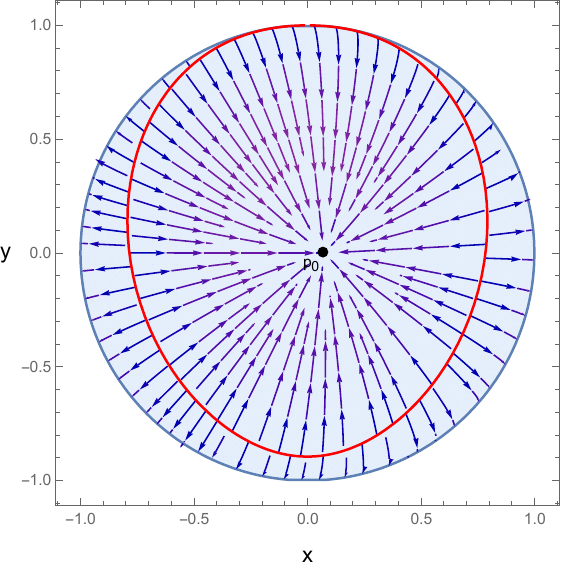}{\bf b)}
\caption{\small The flows of the system (\ref{EqsonCyl0012n})-(\ref{q-disk2}) with fixed $z$ in the disk $\mathbf{D}^2$: {\bf a)} $a^2=0.8$, $z=0.43$,  {\bf b) $a^2=0.8$, $z=0.58$}.  The critical curves are shown by red, while the critical points $p_{0}$ are marked by black.}
\label{Fig: CrPCzPos2}
\end{figure}

In Figs.~\ref{Fig: CrPCzPos2} the positions of $p_{0}$ and flows of the 2d system in the disk are depicted for $a^2=0.8$ and the values of $z$, which correspond to the extrema of the $x$-coordinate of $p_{0}$, i.e. the values of $z$ at which $V''_{\phi}=0$.

\subsection{Compactified $3d$ autonomous system and its critical sets}
In this subsection, we discuss the 3d dynamical system in the unit cylinder and its properties.

Applying the coordinate transformations \eqref{ch-cor} for $X$ and $Y$ to  eqs. (\ref{dz/da})-(\ref{dy/da}), we come to the dynamical system in the $3d$ unit cylinder 
\bea \label{EqsonCyl001}
z' &= &z (z-1) x \, ,\\
 \label{EqsonCyl0012}
       x' &= &\mathrm{p}(x,y,z),  \\  
\label{EqsonCyl0013}
           y'&= &\mathrm{q}(x,y,z) ,
\eea
where $\mathrm{p}$, $\mathrm{q}$ are given by (\ref{q-disk2}),
and we assume (\ref{derred}) for the derivatives. To describe flows of the system (\ref{EqsonCyl001})-(\ref{EqsonCyl0013}), we first discuss  some of its topological  properties.

\begin{figure} [h!]
    \centering
     \includegraphics[width=12cm]{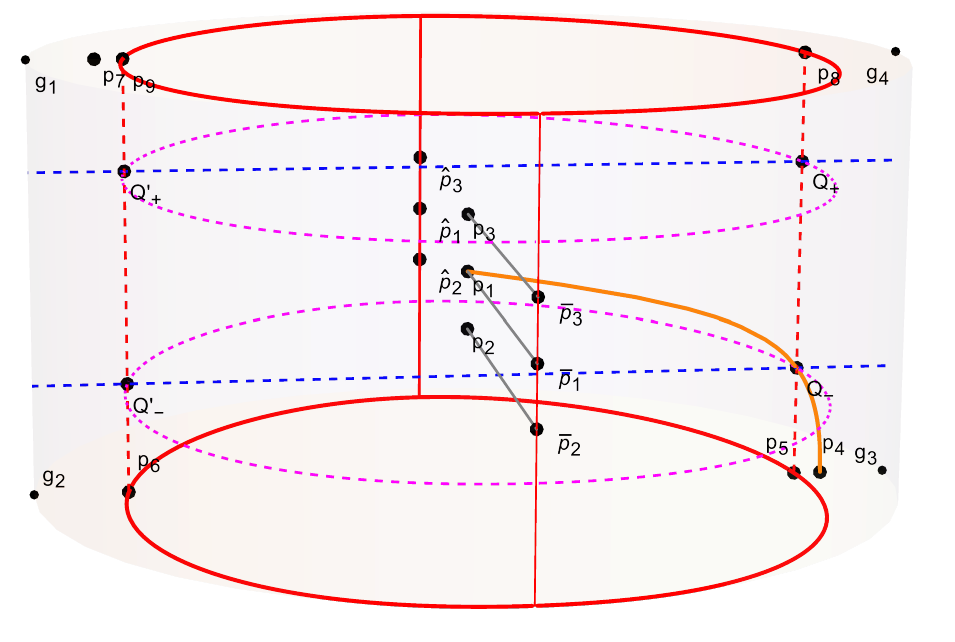}
    \includegraphics[width=3.3cm]{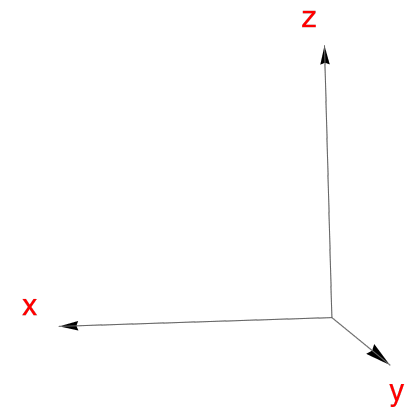}
     \caption{\small The location of the critical sets, regular, singular points and the exact holographic RG flow of the dynamical system (\ref{EqsonCyl001})-(\ref{EqsonCyl0013}) for $a^2=0.8$. The critical  curves 2(a), the lines 2(b),2(d) are shown by red, the closed curves $e_{\pm}$ by magenta (dashed), the lines $N^{\pm}$ are shown by blue (dashed), the lines $M^{\pm}_{x}$ by red (dashed). The exact holographic RG flow $p_{1}-p_4$ is shown by the solid orange  curve. }
     \label{Fig:cylinder}
\end{figure}

The  critical points and sets of eqs.(\ref{EqsonCyl001})-(\ref{EqsonCyl0013}) are:
\begin{enumerate}
\item $xz$-plane 
\begin{enumerate}
\item $p_1$: $(0,0,z_1)$; $p_2$: $(0,0, z_2)$; $p_3$: $(0,0,z_3)$
\item $p_4$: $(-2a^2/\sqrt{4a^4+1},0,0)$; $p_{5,6}$: $(\mp\sqrt{2} a/\sqrt{2a^2+1},0,0)$ 
\item $p_7$: $(2a^2/\sqrt{4a^4+1},0,1)$; $p_{8,9}$: $(\mp\sqrt{2}a/\sqrt{2a^2+1},0,1)$ 
\item $q_1, q_2$: $(-1/\sqrt{5},0,0), (1/\sqrt{5},0,1)$ 
\item $g_{2,3}$, $g_{1,4}$: $(\pm1,0,0)$, $(\pm1,0,1)$
\end{enumerate}
\item $xyz$-space
\begin{enumerate}
\item the closed curves $e_1, e_2$: $z=0,1$;  $x,y \in e(a)$
\item the line $l$: $(0,1,z)$, $z\in[0,1]$
\item the special points on line $l$: $\bar{p}_1, \bar{p}_{2,3}=p_1,p_{2,3} +(0,1,0)$
\item the line $\hat{l}$: $(0,-2/\sqrt{5},z)$, $z\in [0,1]$
\item the special points on the line $\hat{l}$: $\hat{p}_1, \hat{p}_{2,3}=p_1,p_{2,3} +(0,-2/\sqrt{5},0)$.
\end{enumerate}
\end{enumerate}
 The values of $z_1,z_{2,3}$ for the coordinates of the critical points $p_{1},p_{2,3}$, $\bar{p}_{1},{p}_{2,3}$, $\hat{p}_{1},\hat{p}_{2,3}$ are given by  \eqref{ext-in-z} and are related with the extrema of the scalar potential. So for $a^2\leq \frac{1}{2}$, the points $p_{2,3}$, $\bar{p}_{2,3}$, $\hat{p}_{2,3}$ are absent. The  closed curve $e(a)$ is defined by eq.(\ref{clcurv}).
 The critical points 1(a)-1(c),1(e), the critical sets 2(a)-2(e) in the unit cylinder are shown in Fig.~\ref{Fig:cylinder} for $a^2=0.8$,  since this case is most general. To illustrate their positions we also depict the exact half-supersymmetric RG flow from the work \cite{Deger:2002hv} (see Appendix~\ref{app:AppA}) by the orange solid curve, which connects the points $p_1$ and $p_4$. This RG flow is plotted using $x$ (\ref{ch-cor}) and $z$ \eqref{Zvar} functions, which are calculated, in turn, using the scale factor $A$ \eqref{degerSF} and the scalar field $\phi$ \eqref{degerScF}.

The $xz$-plane is an invariant manifold of the 3d autonomous system (\ref{EqsonCyl001})-(\ref{EqsonCyl0013}). This plane is related with the zero-temperature holographic RG flows and was studied in the works \cite{Golubtsova:2022hfk, NE-RG}. As it was mentioned above the critical points $p_{2,3}$  exist if and only if $1/2<a^2<1$ (which  also applies for additional points $\bar{p} _{2,3}$,$\hat{p}_{2,3}$ appearing for  $Y\neq 0$), since they correspond to the additional extrema of the scalar potential at $\phi_{2,3}$. The reconstruction of the metric near $p_{1,2,3}$ shows that these points correspond to  AdS$_{3}$ spacetimes with different radii. We suppose that the dual field theories related to them are  CFT$_1$, CFT$_2$, CFT$_{3}$, correspondingly. 

Near points $p_{4}$, $p_{5,6}$, $p_{7}$, $p_{8,9}$ the gravity solutions have scaling geometries with a divergent scalar field. The dual field theories possess the scaling invariance but not the conformal one. In what follows we omit the discussion related to points $p_{7}$, $p_{8,9}$, since the scalar field tends to $-\infty$ for them and the holographic dictionary should be completed.
The exact holographic RG flow  lies in the $xz$-plane and exists for all values of $a^2$. For $0<a^2<\frac{1}{2}$ and $\frac{1}{2}<a^2<1$ it runs from $p_{1}$ (CFT$_{1}$ UV) to $p_{4}$ (IR). 
For the special case $a^2=1/2$ the points $p_4,p_7$ merge with $p_{5},p_{9}$, while the points $q_1,q_2$, on the contrary, appear only for this case due to the behavior of $\mathcal{C}_{(z,a)}$ at $z=0,1$. So for $a^2=1/2$ the exact flow goes from $p_{1}$ (UV) to $p_{5}$ (IR) \cite{NE-RG}. The analysis from \cite{Golubtsova:2022hfk,NE-RG}  shows that for the RG flows at zero temperature with the Dirichlet boundary conditions are triggered by a source of the relevant operator, CFT$_1$ ($p_1$) can be interpreted as the UV fixed point and  CFT$_{2,3}$ ($p_{2,3}$) are IR fixed points.

The classification of the critical points on the $xz$-plane from the point of view of the 3d system is presented in Table \ref{table: pointsclass} in Appendix \ref{app:AppC}. Note, that the points $p_{5,6}$ and $p_{8,9}$ are no longer isolated critical points in the sense of the 3d system, since they belong to the critical curves $e_1$ and $e_2$, respectively. The points $g_{1,2,3,4}$ occur due to the Poincar\'e transformation (see the discussion below eqs. (\ref{clcurv})-(\ref{pointDisk})) and were not considered in the works \cite{Golubtsova:2022hfk, NE-RG}. In Appendix \ref{app:AppC2} we briefly list possible  holographic RG flows on the $xz$-plane with different boundary conditions.

Another invariant manifold of our $3d$ system (\ref{EqsonCyl001})-(\ref{EqsonCyl0013}) is given by 
\be \lb{IMxy}
M_{xy}: (x,y) \in e(a) \,, z\in [0,1] \, ,
\ee
for which we have no dynamics on $x,y$.

For the case $1/2<a^2<1$ the system also possesses two additional sets
\be \lb{sets}
NP_{\pm}: (x,y,z_{\pm})\,, (x,y) \in [-1,1] \,,
\ee
where the vector field is singular, however, the points $e_{\pm}$ arising from the intersections of $M_{xy}$ and $NP_{\pm}$
\be\lb{reg3d}
e_{\pm}: M_{xy} \cap NP_{\pm}
\ee
are regular. 

The presence of regular points, where the vector field seems to be  singular, also occurs for the 2d dynamical system on the $xz$-plane, i.e. for $y=0$. 
Let us illustrate what we mean by introducing the following sets
\begin{enumerate}
\item lines $M^{\pm}_{x}$: $(\pm\sqrt{2}a/\sqrt{2a^2+1},0,z),$ $z\in[0,1]$,
\item lines $N^{\pm}$: $(x,0,z_{\pm}),$ $x\in[-1,1]$,
\item points $Q_{\pm},Q'_{\pm}$: $Q'_{\pm}= N^{\pm} \cap M^+_{x}, Q_{\pm}=N^{\pm} \cap M^-_{x}$.
\end{enumerate}
These sets are just a contraction of the sets (\ref{IMxy})-(\ref{reg3d}) to the $xz$-plane. 
For the dynamical system on the $xz$-plane the lines $M^{\pm}_{x}$ are invariant manifolds (there is no dynamics on $x$), the lines $N^{\pm}$ are singular; but as it  was noted in \cite{NE-RG} the points $Q_{\pm},Q'_{\pm}$ are regular,
however, the behavior of the flow near them requires some additional study.

It worth to be noted that the line $l$ from 2(b) corresponds to the case $Y\to \infty$. Moreover, in what follows, we discuss the special points of the line $l$, i.e. 2(c), as near-horizon regions of black hole solutions.

We should make some comments about the points $\bar{p}_{1,2,3}$ and $\hat{p}_{1,2,3}$. We do not provide information about their stability in Table \ref{table: pointsclass}. Since from the point of view of the initial system \eqref{dz/da}-\eqref{dy/da}, the points $\bar{p}_{1,2,3}$ are not critical and, generally speaking, are not strictly defined, while the points\footnote{In this work, we are not at all interested in the region with $Y<0$.} $\hat{p}_{1,2,3}$ lie on the critical line $\hat{l}$. Thus, both $\bar{p}_{1,2,3}$ and $\hat{p}_{1,2,3}$ are not isolated and not classified by standard methods through the Jacobian matrix of the linearized system. However, for the system in the cylinder \eqref{EqsonCyl001}-\eqref{EqsonCyl0013} the points $\bar{p}_{1,2,3}$ exist, but they are not isolated. To see the difference between $\bar{p}_{1,2,3}$ we describe flows starting near these points. Below we also carry out a brief analysis considering slices of the system \eqref{EqsonCyl001}-\eqref{EqsonCyl0013} near $\bar{p}_{1,2,3}$.

\subsection{Flows of the 3d autonomous system in the unit cylinder}

In this subsection, we explore the behavior of the 3d system (\ref{EqsonCyl001})-(\ref{EqsonCyl0013}) in the unit cylinder for different values of the parameter $a^2$. We probe the system looking numerically trajectories of eqs.(\ref{EqsonCyl001})-(\ref{EqsonCyl0013}) with certain initial conditions. In terms of coordinates in the unit cylinder (\ref{Zvar}),(\ref{ch-cor})
 near-horizon regions of black hole solutions are associated  with points on the critical line $l$ (see 2(b) from the list of the previous subsection), where $x=0$, $y=1$, $z\in [0,1]$. The zero-temperature asymptotics corresponds to  the $xz$-plane with  $y=0$. We choose the initial conditions so that the solutions start from a near-horizon region of black holes.

\subsubsection{The case $a^2< 1/2$}

\begin{figure}[h!]
  \centering
\includegraphics[width= 13cm]{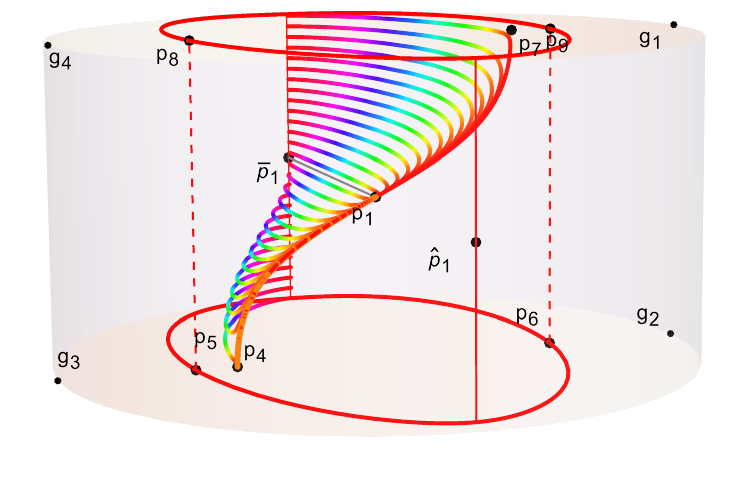}
\caption{\small The  trajectories of the dynamical system in the unit cylinder (\ref{EqsonCyl001})-(\ref{EqsonCyl0013}) calculated numerically for $a^2=0.25$ 
with the initial conditions (\ref{incon1}) for various $\delta$ (shown by colorful curves); the AdS black hole solutions $\bar{p}_1-p_1$ (shown by gray); the exact flows zero-temperature flows $p_1-p_4$ and $p_1-p_7$, which are separatrices of the system (shown by orange curves).}
\label{Fig: CrPCzPos2-025}
\end{figure}

Consider the  case $a^2< 1/2$, for which the scalar potential has only one extremum at $\phi=0$ or in terms of the $z$-variable at $z_{1}$.
To solve eqs.(\ref{EqsonCyl001})-(\ref{EqsonCyl0013}) numerically, we impose the following initial conditions for the equations
\be\label{incon1}
x=0,\;\;\; y=1-\varepsilon,\;\;\; z\in[z_1-\delta,z_1+\delta],
\ee
where $\varepsilon,\delta>0$. For the initial conditions (\ref{incon1}) and those we discuss below, $z$ is actually $z_{h}$, that in turn determines the value of the scalar field at the horizon $\phi_h$  (\ref{Zvar}).The initial conditions (\ref{incon1}) yield two cases of trajectories, namely, the trajectories start at: 1) $z=z_{1}$, 2) $z<z_{1}$ or $z>z_{1}$. Numerical solutions to eqs.(\ref{EqsonCyl001})-(\ref{EqsonCyl0013}) with (\ref{incon1}) for $a^2=0.25$ are depicted in Fig.~\ref{Fig: CrPCzPos2-025}. The solutions begin from the region near the line $l$ ($x=0$, $y=1-\varepsilon$ and $\varepsilon$ is small), i.e. they start from the near-horizon region, then go to the $xz$-plane with $y=0$, which corresponds to zero temperature.

The trajectory for which $\delta=0$ in (\ref{incon1}), i.e. the initial conditions read as $(0,y-\varepsilon,z_{1})$ matches with the AdS black hole solution (\ref{ck-sol}). This trajectory is depicted by the gray line $\bar{p}_{1}-p_{1}$ in Fig.~\ref{Fig: CrPCzPos2-025} starting from the near-horizon region at $\bar{p}_{1}$ and ending at $p_{1}$, which corresponds to the AdS asymptotics. Note, that the trajectory exists for any value of $a^2$,  since it is related with the extrema of the scalar potential at $z_{1}$.

If the initial conditions are set as $(0,y-\varepsilon,z_{1}\pm \delta)$, where $\delta$ is arbitrary, we obtain solutions shown by colorful curves in Fig.~\ref{Fig: CrPCzPos2-025}. For the trajectories with the starting points  above $\bar{p}_{1}$, we observe that they first go to the region with $x>0$ and slowly decreasing $z$, then they tend to $x=0$, while $z$ begins to decrease faster and, finally, the trajectories end at the stable node $p_{1}$ on the $xz$-plane. The similar behaviour occurs for the solutions with a starting point below $\bar{p}_{1}$ with the difference, that the trajectories run into the region with $x<0$ and slowly increasing $z$, which starts to increase faster tending again to the stable node $p_{1}$ at the $xz$-plane. It is important to notice that all such trajectories tending to $p_{1}$ pass close to the exact holographic RG flow ($p_{1}-p_{4}$) and its mirror image (the curve $p_{1}-p_{7}$). So the exact trajectories $p_{1}-p_{4}$ and $p_{1}-p_{7}$ are separatrices of the dynamical system (\ref{EqsonCyl001})-(\ref{EqsonCyl0013}). In Fig.~\ref{Fig: CrPCzPos2-025} they are shown by orange curves (solid and dashed, correspondingly).

\begin{figure}[t]
    \centering
  \includegraphics[width=11cm]{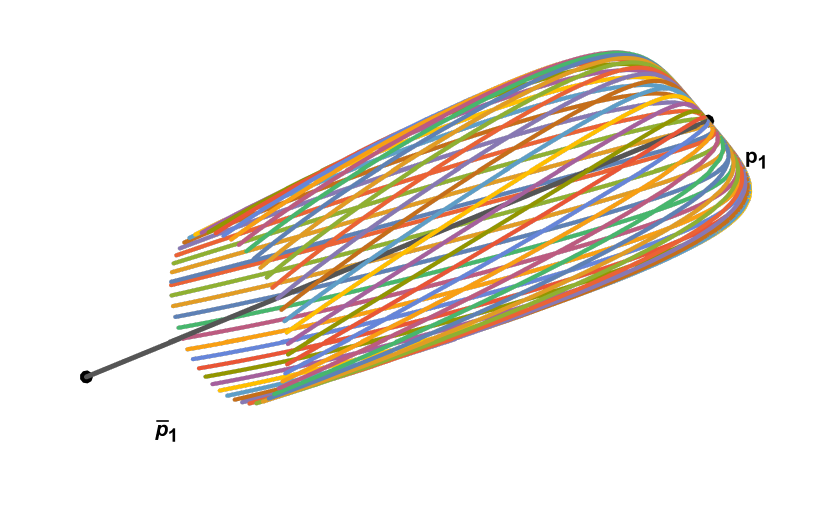}
    \caption{The numerical trajectories of the dynamical system (\ref{EqsonCyl001})-(\ref{EqsonCyl0013}) inside the invariant manifold $M_{xy}$  (\ref{IMxy}) for $a^2=0.25$  with the initial conditions (\ref{inconinv}). For the initial conditions we fix $\rho=0.1$, $\delta=0.5$ and $\varepsilon = 0.2$. The exact AdS black hole solution $\bar{p}_{1}-p_{1}$ is shown by gray.}
    \label{Fig: CrPCzPos4}
\end{figure}

Another important feature of the system is related to the dynamics inside the invariant manifold $M_{xy}$ (\ref{IMxy}), namely, all trajectories inside $M_{xy}$ have a similar structure. To illustrate this, it is convenient to impose the following initial conditions on eqs.(\ref{EqsonCyl001})-(\ref{EqsonCyl0013}):
\be\label{inconinv}
x=\rho\cos\psi,\;\;\; y=1-\varepsilon,\;\;\; z\in\rho\sin\psi+\delta,
\ee
with sufficiently small $\rho,\delta$ and $\psi\in[0,2\pi]$.
Then the trajectories start from a point on the circle  with the center in  $\bar{p}_{1}$ and the radius $\rho$, generally coinciding with the behaviour of the  trajectories from the previous item and coming to the stable node $p_{1}$. The solutions constructed numerically with (\ref{inconinv}) are shown in Fig.~\ref{Fig: CrPCzPos4}. 

\subsubsection{The case $1/2<a^2<1$}

Now we turn to the discussion of the case $1/2<a^2<1$,  for which the potential has three extrema at $\phi_{1,2,3}$ or in terms of $z$  at $z_{1,2,3}$.

\begin{figure} [t]
    \centering
     \includegraphics[width=13cm]
     {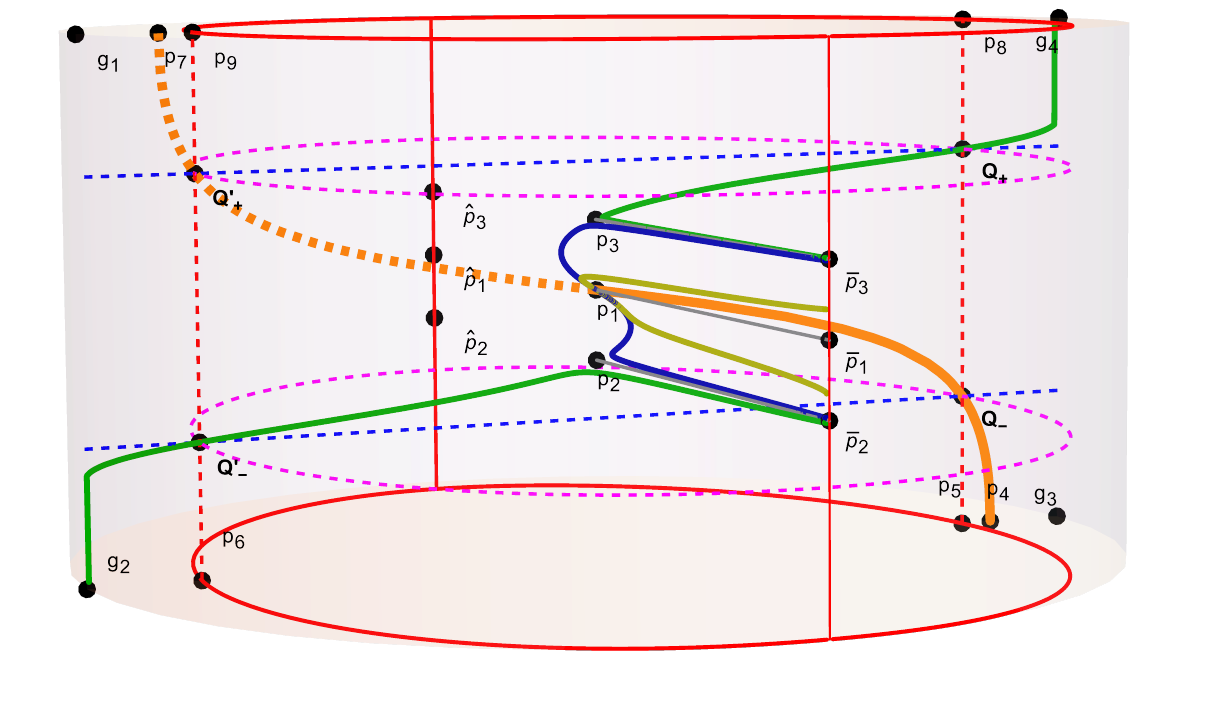}
     \caption{\small  Numerical solutions, exact solutions, critical sets, regular and singular points of the $3d$ dynamical system (\ref{EqsonCyl001})-(\ref{EqsonCyl0013}) with  $a^2={0.8}$ in the unit cylinder. Numerical trajectories  are: $\bar{p}_{1,2,3}-p_{1,2,3}$ (shown by gray), $\bar{p}_{2,3}-p_{1}$ (shown by blue), $\bar{p}_{2}-g_{2}$, $\bar{p}_{3}-g_{4}$ (both shown by green), numerical solutions from 2(b) (both shown by olive). The exact solutions $p_{1}-p_{4}$ and its mirror image $p_{1}-p_{7}$ are shown by thick orange curves (solid and dashed, correspondingly).}
     \label{Fig:cylinder-case-one}
\end{figure}

To find solutions of our interest to eqs. 
(\ref{EqsonCyl001})-(\ref{EqsonCyl0013}), which start again from the critical line $l$,  we suppose the following class of the initial conditions:
\be \lb{icC2}
x=0,\;\;\; y=1-\varepsilon,\;\;\; z\in[z_2-\delta,z_3+\delta]\footnote{We recall that the points $z$ corresponding to the extrema of the scalar potential have the following order $z_{2}<z_{1}<z_{3}$.},
\ee 
where $\varepsilon, \delta>0$ are small.
  These initial conditions can be naturally divided into three types: \begin{enumerate}
  \item $z=z_{1,2,3}$, \item $z_2 < z < z_3$, \item $z<z_2$ or $z>z_3$.\end{enumerate}
In Fig. \ref{Fig:cylinder-case-one} we depict numerical trajectories of eqs. (\ref{EqsonCyl001})-(\ref{EqsonCyl0013}) with the initial conditions \eqref{icC2} for $a^2=0.8$ for all cases of starting points. 
These solutions begin from the region near the line $l$, with $y=1-\varepsilon$, and, then, tend to the $xz$-plane with $y=0$, i.e. to the zero-temperature plane, however, otherwise they behave differently. 
Below we give the classification of these flows in details.
\begin{enumerate}
    \item Fixing the initial conditions for the system as: $(0,y-\varepsilon, z_{1,2,3})$, i.e.  $\delta=0$, we obtain three solutions coinciding with the exact AdS black holes (\ref{ck-sol}) (see also Appendix \ref{app:AppBM}). 
    From the point of view of the phase diagram the trajectories are lines, that connect the points $\bar{p}_{1,2,3}$ with the points $p_{1,2,3}$, respectively. We show the trajectories $\bar{p}_{1,2,3}-p_{1,2,3}$ by gray color in Fig. \ref{Fig:cylinder-case-one}. Note, that we have three different AdS black hole solutions corresponding to the extrema of the potential. Moreover, types of stability of the ending points $p_{1,2,3}$ are different, see Table \ref{table: pointsclass}. So the solution $\bar{p}_{1}-p_{1}$ is related with the thermal state of UV CFT$_1$, while 
     $\bar{p}_{2}-p_{2}$ and $\bar{p}_{3}-p_{3}$ are related with the thermal states of IR CFT$_2$ and CFT$_3$, respectively.
    \item 
    \begin{enumerate}

     \item Let us fix the initial conditions as: $(0,y-\varepsilon, z_{1}\pm \delta)$, where $\delta$ is not necessarily small,  so, that the starting point on the $z$ axis should not be below $z_2$ or above $z_3$. These solutions are shown by olive in Fig. \ref{Fig:cylinder-case-one}.
    Note, that if we take $\delta \to 0$, then such solution is almost identical to the  solution $\bar{p}_1- p_1$ from item {\bf 1}.  The trajectories behave as follows. Setting the initial point along the $z$-axis below the point $\bar{p}_1$, the trajectory first tends to the region with a negative $x$, while $z$  slowly increases, then, the trajectory runs to $x=0$, and $z$ increases faster, i.e. the closer the solutions are to the $xz$-plane, the faster $z$ grows, and eventually the trajectory ends at the stable node $p_1$.
    One can give a similar description for a trajectory that begins above $\bar{p}_1$. The difference is that this trajectory tends to the region with positive $x$, while $z$ slowly decreases, then, the solution goes to $x=0$, and $z$ decreases faster, so the solution reaches the stable critical point $p_1$.
    
    \item 
    Now we set the initial conditions as: $(0,y-\varepsilon, z_{2,3}\pm \delta)$. As a result, we get two trajectories, which connect $\bar{p}_{2,3}$ and $p_1$ 
    passing the unstable (saddles) critical points $p_{2,3}$ with $y=0$. We depict these flows by dark blue color in Fig. \ref{Fig:cylinder-case-one}. In this case there is some peculiarity, namely, if one sends  $\delta \to 0$, the flow can be conditionally split into two parts, i.e. $\bar{p}_{2,3}-p_{2,3}$ and $p_{2,3}-p_1$. 
   Moreover, the smaller $\delta$ , the closer the first part of the flow is to the exact AdS black hole solution (connecting $\bar{p}_{2,3}$ and $p_{2,3}$), which was described above; and the second part is closer to the flow between two pure AdS solutions with different radii, $p_{2,3}-p_{1}$, which have been considered in the works \cite{Golubtsova:2022hfk, NE-RG}.
    
  \end{enumerate}
  \item Fixing the initial conditions as: $(0,y-\varepsilon, z_{2,3}\mp \delta)$ with small $\delta$, we find the solutions from $\bar{p}_{2,3}$ to the stable critical points $g_{2,3}$. These solutions can be interpreted as  naked singularities. We depict the trajectories by green color in Fig. \ref{Fig:cylinder-case-one}. Let us discuss a solution, which starts at $(0,y-\varepsilon, z_2-\delta)$ (a similar description can be given for the second flow, which starts near $z_3$). For small $\delta$, the solution is splitted into two parts:  $\bar{p}_2-p_2$ and  $p_2- g_2$. Again, the smaller $\delta$, the more the first part coincides with the  AdS black hole solution ($\bar{p}_2-p_2$), described above in the first item. The part of the solution $p_2-g_2$ lies in the invariant $xz$-plane, passing through the regular point $Q'_{-}$. It worth to be noted, that the points $g_{2,3}$ correspond to scaling singular geometries, which do not satisfy Gubser's bound, so these flows are not well defined from the point of view of holography. 
\end{enumerate}

The behavior of the solutions  with \eqref{icC2} is always exactly the same as we presented above. This can be understood from the following observation: we determine the dynamics of the trajectory in the $xy$-plane using by the position of the critical point $p_0$ with the coordinates given by (\ref{pointDisk}), which in turn depend on $z$. Then the dynamics along the $z$-axis is determined directly from the simple equation (\ref{EqsonCyl001}) and mostly depends on a sign of $x$. We should also take into account that the motion in the $xy$-plane affects the motion along the $z$-axis and vice versa.

\begin{figure}[t]
    \centering
  \includegraphics[width=11cm]{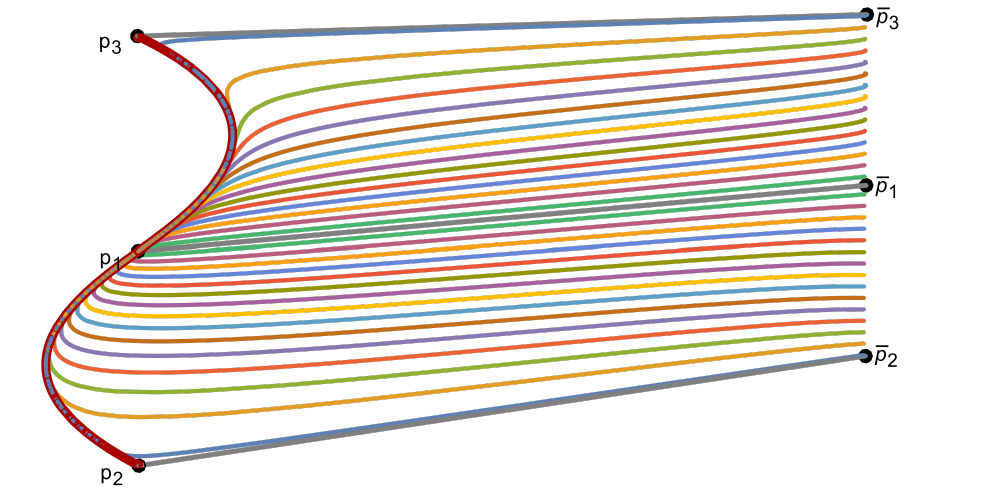}
    \caption{The numerical trajectories of the dynamical system (\ref{EqsonCyl001})-(\ref{EqsonCyl0013}) for $a^2=0.8$  with the initial conditions (\ref{icC2}) and the separatrices $p_{1}-p_{2}$, $p_{2}-p_{3}$.}
    \label{Fig: SyclineAdS2}
\end{figure}

Consider the trajectory from item 2(b). For the  initial condition (\ref{icC2}), we choose the value of $z$ such that $z_2<z<z_1$ and $z<(z_1-z_2)/2$, i.e. it lies closer to $z_2$ than to $z_1$. The solution runs from the region below the point $\bar{p}_{1}$ with slowly increasing $z$. To understand the dynamics of the  trajectory in the $xy$-plane, we need to take a look at Section \ref{pdis}. The point $p_0$ in this case is a stable node (see  Table \ref{table:p02D-t}), which is located  in the negative region of $x$ (see the green curve in Fig. \ref{Fig: CrPCzPos} \textbf{b)}). Therefore, the flow tends to $p_{0}$,  decreasing in both $x$ and $y$. As far as $x<0$,   $z$ grows, that follows from eq. (\ref{EqsonCyl001}). Next, since $z$ increases, then from Fig. \ref{Fig: CrPCzPos} \textbf{b)} it follows that the position of the point $p_0$ goes further and further into the negative region of $x$ until $z$ passes the local minimum of the function $p_0(z,a)$, which is located between $z_2$ and $z_1$  and corresponds to $V''_{\phi}=0$ (see  Fig. \ref{Fig: CrPCzPos} \textbf{b)} again). After passing this minimum, the position of $p_0$ along $x$ tends to $x=0$, but at the same time, since $x$ is still negative, $z$ grows. Finally, the trajectory ends at the stable node $p_1$ with $x=0$.

It is important to note that  black hole solutions with the initial conditions
 (\ref{icC2}) such that $z_{2}<z<z_{1}$ or $z_{1}<z<z_{3}$, i.e. the values of the scalar field on the horizon $\phi_{h}$ takes a value between extrema of the scalar potential, pass close to the trajectories $p_{1}-p_{2}$ and $p_{1}-p_{3}$, which are zero-temperature holographic RG flows between two AdS fixed points \cite{NE-RG}. Thus, the flows $p_{1}-p_{2}$ and $p_{1}-p_{3}$ are separatrices in this case. We depict these solutions in Fig.\ref{Fig: SyclineAdS2}.
 
One can consider a more general class of initial conditions comparing to \eqref{icC2}\footnote{For the readability we have split the discussion of the initial conditions (\ref{icC2}) and (\ref{in-con-nnx})}, namely
\be \lb{in-con-nnx}
x=0\pm \upsilon, \quad y=1-\varepsilon, \quad z\in [z_2-\delta,z_{3}+\delta],
\ee
i.e. we add a variation by $x$ with small $\varepsilon,\upsilon>0$ such that $\upsilon$ and $\varepsilon$ respect the constraint (\ref{cyl-1}).
Thus, the initial conditions are divided into the following cases:
\begin{enumerate}
    \item  $x=\pm \upsilon,\quad y=1-\varepsilon,\quad z=z_{1,2,3}$.
    \begin{enumerate}
        \item For the initial conditions $(- \upsilon,y-\varepsilon,z_{1})$ the solution starts from the region near $\bar{p}_{1}$ with negative $x$. First, $y$ decreases and $z$ increases (see eq. \ref{EqsonCyl001}), such that in the $xy$-plane the flow tends to the attracting point $p_{0}$ (a stable node, see Table \ref{table:p02D-t}).  The coordinate $x$  starts to increase  being negative, so $p_{0}$ tends into the region with $x>0$, see  Fig. \ref{Fig: VecFieInSys2d} and Figs. \ref{Fig: CrPCzPos2} {\bf a)} and {\bf b)}. Since $x$ is still negative, $z$  increases but more and more slowly.
        Next,  $x$ reaches $0$, $p_{0}$ turns and goes to the direction with  $x<0$,  while $z$ starts to decrease. So the trajectory ends at the stable $p_{1}$.
        
        \item The trajectory with the initial conditions $x=\upsilon$, $y=1-\varepsilon$, $z=z_{1}$ begins near $\bar{p}_{1}$ with a positive $x$. Here, since $x>0$, $z$ decreases as well as $y$ and the trajectory tends again to the stable node $p_{0}$ in the $xy$-plane. The point $p_{0}$ goes to the region with  $x<0$. The coordinate $z$  decreases slowly, and begins to increase when $x$ reaches $0$. Finally, the trajectory ends at $p_{1}$.
        \item Imposing the initial conditions $x=0-\upsilon$, $y=1-\varepsilon$, $z=z_{2}$ and $x=0+\upsilon, y=1-\varepsilon$, $z=z_{3}$ on eqs. (\ref{EqsonCyl001})-(\ref{EqsonCyl0013}) gives rise the  trajectories, which are similar to $\bar{p}_{2}-p_{2}-p_{1}$ and $\bar{p}_{3}-p_{3}-p_{1}$, correspondingly. This behaviour was described in item 2(a) of the previous list.
        \item Changing the sing of $\upsilon$ in the conditions from the previous item yields the trajectories that start from $\bar{p}_{2,3}$, then pass through $Q^{'}_{-}$ and $Q_{+}$, respectively and go $z=0,1$, which correspond to  $\phi\to \pm \infty$.
    \end{enumerate}
    Note, that the flows with the initial conditions (\ref{icC2}) considered above behave monotonically on $z$, while for the trajectories with the initial conditions $x=0\pm \upsilon$, $y=1-\varepsilon$, $z=z_{1}$ the behaviour on $z$ is non-monotonic.
        \item  The flows with the initial conditions $x= 0 \pm \upsilon$, $y=1-\varepsilon$, $z=z_{1}\pm \delta$ can have both monotonic and non-monotonic behaviours with respect to $z$, but finally end at the same stable point $p_1$. For them, a picture similar to Fig. \ref{Fig: CrPCzPos4} is valid.  In the next section we will present analytical formulae to describe these trajectories near the point $\bar{p}_{1}$.
        
       \item Assuming the initial conditions of the type: $x=\pm \upsilon$,$y=1-\varepsilon$, $z=z_{2,3}\pm \delta$, we construct the flows, which do not necessarily end at the stable critical point $p_1$, in contrast to the trajectories from the previous item. To illustrate  the most significant feature of such flows, let us describe a trajectory starting near the fixed point $\bar{p}_3$ (the case of the flows from $\bar{p}_2$ can be described similarly). Consider, for example, the initial condition of the form $(-\upsilon, 1-\varepsilon, z_3-\delta)$. First, recall that if $\upsilon=0$ we would get a flow from  $\bar{p}_{3}$ running to the point $p_1$ passing $p_3$ (see Fig. \ref{Fig:cylinder-case-one}). However, a suitable variation of the initial conditions along $x$ in the negative direction can lead to a flow going directly to the stable point $g_4$ passing the unstable point $p_3$, see Table \ref{table: pointsclass}. A similar picture is observed for the initial conditions $x=\upsilon$, $y= 1 -\varepsilon$, $z=z_3+\delta$. In this case, with a certain choice of $\upsilon$, the flow can go to $p_1$, however as it has already been described  above (see Fig. \ref{Fig:cylinder-case-one}),  for $\upsilon=0$ the flow definitely goes to $g_4$.  
       \end{enumerate}

\begin{figure} [h!]
    \centering
     \includegraphics[width=7.5cm] {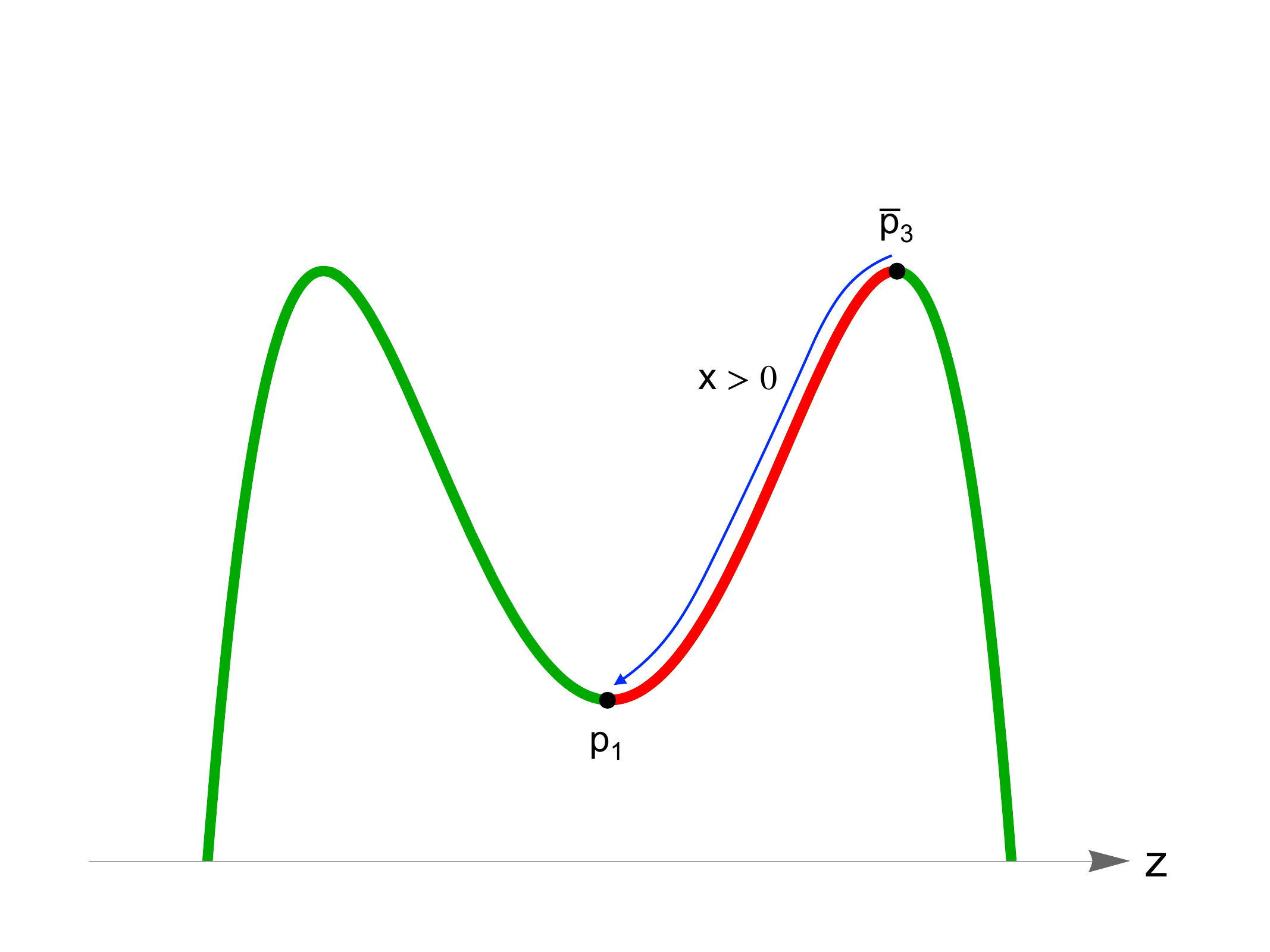} {\bf a)}
    \includegraphics[width=7.5cm] {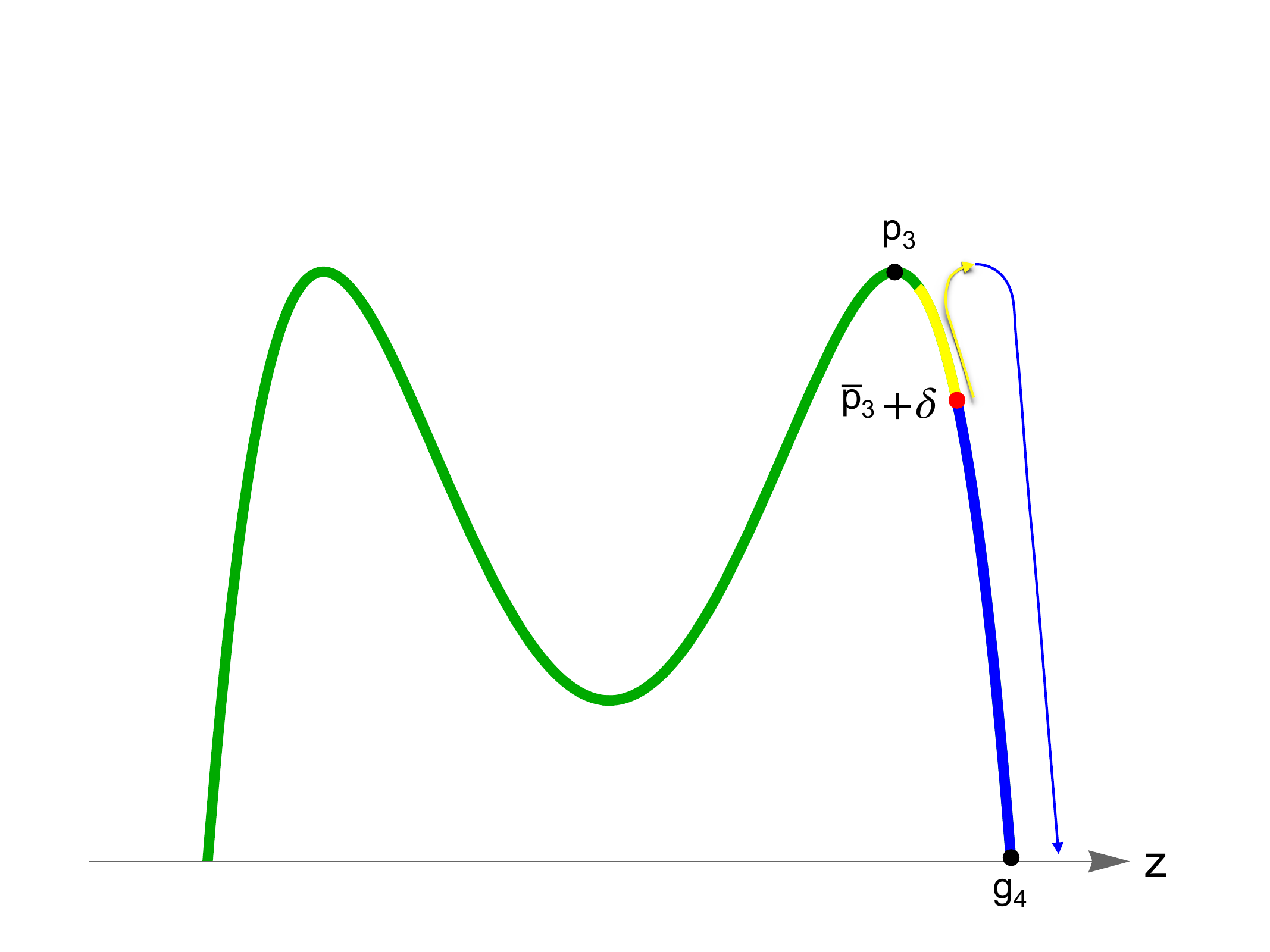} {\bf  b)}
     \caption{\small The motion of the point particle with different initial conditions. a) The trajectory begins at the point $\bar{p}_3$ with $x = \upsilon$, such $\upsilon> 0$, and ends at the point $p_1$.  The red stripe in the figure shows the trajectory of the particle, the blue arrow indicates the direction of motion. b) The trajectory begins at  $\bar{p}_3 + \delta$, with $x=\upsilon$ where $\upsilon>0$ is sufficiently small. The trajectory first  moves towards the point $p_3$, this part of the trajectory is depicted by yellow, while the direction of motion is marked by the yellow arrow. At certain moment, $x$ becomes negative and $\dot{z}>0$. As a result, the trajectory moves to $g_4$, this part of the trajectory is shown by  blue,  the direction of motion is shown by the blue arrow.}
     \label{fig:mechmodpic}
\end{figure}

       In fact, the behavior of the trajectories specified by the initial conditions \eqref{in-con-nnx} can be described as follows. Let us consider the motion of a one-dimensional point  particle in the potential with a minimum at the origin and two symmetric maxima to the left and right of the origin, i.e.  the  potential, which is inverse to that one in Fig. \ref{fig:scpotential} with $a^2=0.8$. Let the coordinate of the point particle be denoted as $\mathsf{x(A)}$, and the velocity as $\dot{\mathsf{x}}(A)$, where $A$ is time. It is not difficult to verify that the functions $z(A)$ and $-x(A)$ in the cylinder behave like the quantities $\mathsf{x}(A)$ and $\dot{\mathsf{x}}(A)$, respectively.  In Fig.\ref{fig:mechmodpic} we show the possible dynamics of the particle. In Fig.\ref{fig:mechmodpic} {a)} we depict the trajectory of the point particle, which begins at the point $\bar{p}_{3}$ with the velocity $-\dot{\mathsf{x}}(A)=\upsilon$ such that $\upsilon>0$ and ends at the minimum $p_{1}$. This trajectory corresponds to the flow from the list 1(c) with (\ref{in-con-nnx}). In Fig. \ref{fig:mechmodpic} {b)} we show the case, when the particle starts its motion from $\bar{p}_{3}+\delta$ towards $p_{3}$ with a quite small velocity $-\dot{\mathsf{x}}(A)=\upsilon$, $\upsilon>0$. Before reaching the point $p_3$, the particle changes the direction of motion and ends the trajectory at the point $g_4$.

       In the next section we will give an analytical description for the flows near the critical line $l$ described above.

 \setcounter{equation}{0}
 
 \section{Near-horizon solutions from the dynamical system}\label{sec:nhsol}
 
  In this section, we reconstruct  asymptotic finite temperature solutions near the fixed points. The  solutions can be associated with near-horizon geometries of asymptotically AdS black holes. We also briefly discuss properties of the fixed points on the line $l$.
 
 \subsection{Qualitative analysis of critical points at finite temperature}
 In this subsection we give a qualitative analysis of fixed points of the $3d$ system (\ref{EqsonCyl001})-(\ref{EqsonCyl0013}) in the cylinder, which are related with non-zero temperature.
Let us consider a region of our $3d$ system (\ref{EqsonCyl001})-(\ref{EqsonCyl0013}) near the critical line $l$, i.e. under the condition $x \sim 0$ and $y\sim 1$.  It is convenient to rewrite the latter condition introducing a new variable $\zeta$ 
\be \lb{y-e-ch}
y=1-\zeta,
\ee
so that  $\zeta$ is sufficiently small. Taking into account (\ref{y-e-ch}) and keeping only the first order in $x$ and $\zeta$ we represent
\be \lb{SqrA}
\sqrt{1-x^2-y^2} \sim \sqrt{2\zeta} \,. 
\ee
Plugging (\ref{y-e-ch}) and (\ref{SqrA}) into eqs. (\ref{EqsonCyl001})-(\ref{EqsonCyl0013}) and keeping the most significant contributions in $x$ and $\zeta$, we are brought to the following dynamical system
\bea 
\lb{Aeq-z}
\frac{dz}{dA} &=& \frac{ z(z-1) \, x }{\sqrt{2\zeta}} \,, \\
\lb{Aeq-x}
\frac{dx}{dA} &=& - \mathcal{C}_{(z,a)} \,, \\
\lb{Aeq-y}
\frac{d\zeta}{dA} &=& \sqrt{2\zeta} \,,
\eea
where it is convenient to come from the derivatives defined by eq. \eqref{derred} to ordinary ones.

Before solving eqs.\eqref{Aeq-z}-\eqref{Aeq-y}, we will discuss some interesting properties of trajectories of the 2d dynamical system, which is obtained as a slice of the system \eqref{Aeq-z}-\eqref{Aeq-y} along the variable $\zeta$ ($0<\zeta<<1$) . Fixing $\zeta$ the system \eqref{Aeq-z}-\eqref{Aeq-y} reduces to
\bea \lb{r-Aeq}
z' &=& z(z-1)x \,, \\ \lb{r-Aeq2}
x' &=& -\mathcal{C}_{(z,a)} \sqrt{2\zeta},
\eea
where we again use the redefinition of the derivatives \eqref{derred} taking into account \eqref{SqrA} to eliminate the divergence as $\zeta$ is sent to $0$. 
The dynamical system \eqref{r-Aeq}- \eqref{r-Aeq2} on the $xz$-plane at some $\zeta\neq 0$ has the following critical points
\be \lb{c-p-t}
\mathsf{z}_1= (z_1,0) \,, \quad 
\mathsf{z}_2 = (z_2,0) \,, \quad
\mathsf{z}_3 = (z_3,0),
\ee
with the $z$-coordinates corresponding to the extrema of the potential (\ref{ext-in-z}). Thus, for $0<a^2\leq\frac{1}{2}$  we have only one critical point $\mathsf{z}_{1}$, since the points  $\mathsf{z}_{2,3}$ turn to be complex in this case and for $\frac{1}{2}<a^2<1$ there are three critical points. The critical points $\mathsf{z}_{1,2,3}$ can be considered as a projection of the critical points $\bar{p}_{1,2,3}$ onto the $xz$-plane with a non-zero $y$, which is close $1$.

To classify the points $\mathsf{z}_i$ we find the eigenvalues $\lambda^i_{1}$,$\lambda^i_2$ of the Jacobian matrix of the linearized system \eqref{r-Aeq}-\eqref{r-Aeq2}. Omitting the calculations we get
\be \lb{eigen-zx}
\lambda^i_{1,2} = \pm (2\zeta)^{1/4}\left. \left ( \sqrt{|z(z-1)|}  \sqrt{\frac{d \mathcal{C}_{(z,a)}}{d z}} \right ) \right|_{z=z_i},
\ee
where $i=1,2,3$.

\begin{figure} [h!]
    \centering
     \includegraphics[width=6.5cm] {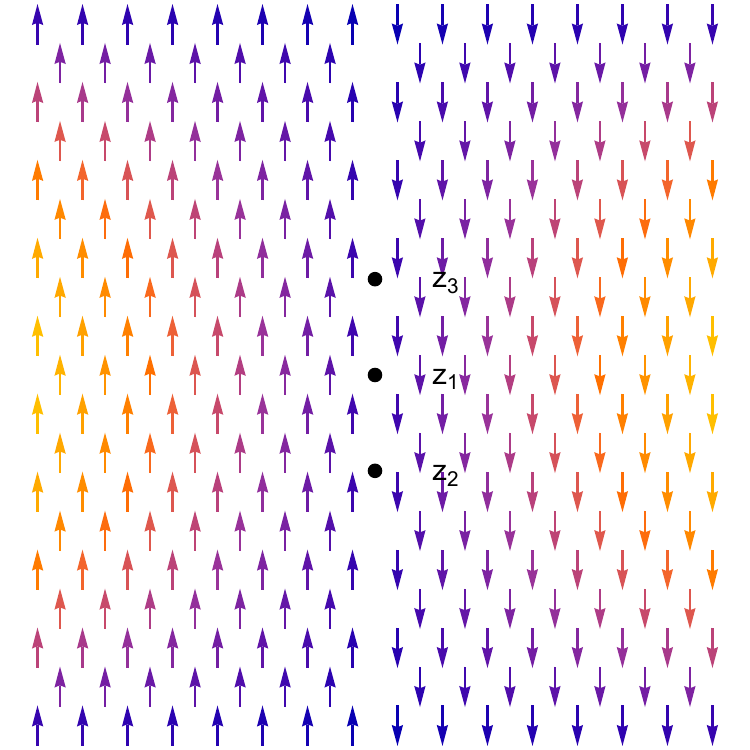} {\bf  a)}
     \includegraphics[width=6.8cm] {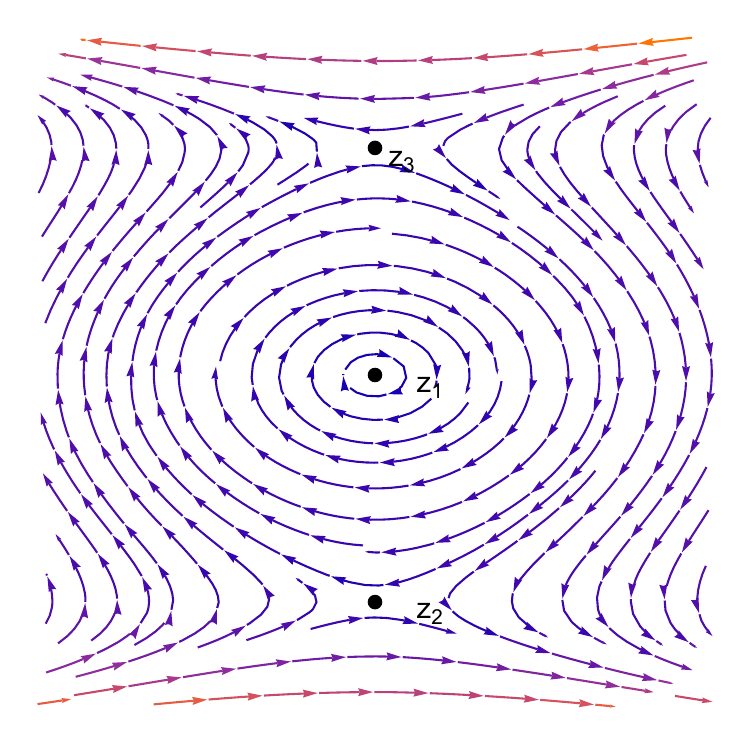} {\bf  b)}
     \caption{\small Phase portraits of the system \eqref{r-Aeq}-\eqref{r-Aeq2} in small neighborhoods of the points $\mathsf{z}_{i}$ for $a^2=0.8$: \bf{a)} $\zeta=0$, \bf{b)} $0<\zeta<<1$.}
     \label{Fig:starpoints2}
\end{figure}

Below we consider the case $\frac{1}{2}<a^2<1$, then  the following expressions   are valid
\be \lb{conVV}
\left.\frac{d \mathcal{C}_{(z,a)}}{d z}\right|_{z=z_1}<0 \,, \quad 
\left.\frac{d \mathcal{C}_{(z,a)}}{d z}\right|_{z=z_{2,3}}>0.
\ee
Thus, if $\zeta\neq0$ the point $\mathsf{z}_1$ is center and the points $\mathsf{z}_{2,3}$ are saddles.  The phase portraits of the system \eqref{r-Aeq}-\eqref{r-Aeq2}  with $a^2=0.8$ are shown in Figs.\ref{Fig:starpoints2} {\bf a)} and {\bf b)} for $\zeta=0$ and $0<\zeta<<1$, correspondingly.
 
The value $\zeta$ in \eqref{r-Aeq}-\eqref{r-Aeq2}  can be considered as a bifurcation parameter of the dynamical system: at $\zeta = 0$, the phase portraits in small neighborhoods of the points $\mathsf{z}_{i}$ look the same (see Fig. \ref{Fig:starpoints2} \textbf{a)}), and as $\zeta$ increases, a center is formed near $\mathsf{z}_1$, and saddles are formed near $\mathsf{z}_{2,3}$ (see Fig. \ref{Fig:starpoints2} \textbf{b})).

This estimation of stability for the critical points $\bar{p}_{1,2,3}$ allows to assume that the critical point $\bar{p}_{1}$ is stable, while  $\bar{p}_{2,3}$ are unstable. However, we note that this estimation should be improved, for instance, by the center manifold analysis \cite{Wigg} as it was done in \cite{Ganguly:2014qia}.

\subsection{Reconstruction of asymptotic solutions near critical points}

In this subsection we derive solutions for the metric functions $A$,$f$ and the scalar field $\phi$ near fixed points at finite temperature, i.e. we construct near-horizon AdS black-hole solutions.

Below we show that near the critical points $\bar{p}_{1,2,3}$ (and in general near the critical line $l$ on which these points lie, see Fig. \ref{Fig:cylinder}) the simplified version of the system \eqref{Aeq-z}-\eqref{Aeq-y} can be solved exactly. Solutions of this simplified system fully reproduce the flows near the $l$ line described in the previous section (see Fig. \ref{Fig: CrPCzPos2} or Fig. \ref{Fig:cylinder-case-one}). However to restore the metric functions $A$ and $f$ we will use the simplest class of these solutions.

As one can see, the latter equation of the system \eqref{Aeq-z}-\eqref{Aeq-y} 
is independent  of other variables and can be solved immediately
\be \lb{Aeq-y-s}
\zeta = \frac{1}{2}(A-A_h)^2,
\ee
where $A_h$ is some positive constant (see Appendix~\ref{app:AppBM}). Taking into account \eqref{Aeq-y-s}, we can rewrite the remaining two equations of (\ref{Aeq-z})-(\ref{Aeq-x}) in the following form
\bea
\lb{Aeq-zx-m}
\frac{dz}{dA}& =& \frac{z(z-1) \, x}{(A-A_h)} \,, \quad\\ \lb{Aeq-zx-m2}
\frac{dx}{dA} &=& - \mathcal{C}_{(z,a)} \,. 
\eea
In terms of the original variables $X$ and $Y$ (see eqs.\eqref{dz/da}-\eqref{dy/da}), the  system \eqref{Aeq-zx-m}-\eqref{Aeq-zx-m2} is represented as
\bea \lb{Aeq-dz-dx}
\frac{dz}{dA} &=& z(z-1) X \,, \quad \\
\lb{Aeq-dz-dx-1}
\frac{dX}{dA} &=& - \frac{1}{(A-A_h)} (X+\mathcal{C}_{(z,a)}) \, , 
\eea
where  the coordinate transformations \eqref{ch-cor} are used. This is in agreement with the near-horizon form of  equations of motion for $X$ and $Y$  from \cite{Gursoy:2008za}.

Note, that eqs. (\ref{Aeq-dz-dx})-\eqref{Aeq-dz-dx-1} can be obtained from the original equations \eqref{dz/da}-\eqref{dy/da} almost immediately if we send $Y \to \infty$ for them. In this case, the equation in $Y$ is also separated and has a solution
\be \lb{forY-nn}
Y(A) = \frac{1}{A-A_h}.
\ee
Plugging eq.\eqref{forY-nn}  back into the remaining ones  \eqref{dz/da}-\eqref{dy/da}, we get exactly eqs.\eqref{Aeq-dz-dx}-\eqref{Aeq-dz-dx-1}.

To rewrite eqs.\eqref{Aeq-dz-dx}-\eqref{Aeq-dz-dx-1} in a simpler form, it is convenient to come from  the variable $z$ back to $\phi$, i.e. to use the transformation \eqref{Zvar}. Then the system \eqref{Aeq-dz-dx}-\eqref{Aeq-dz-dx-1} reads as
\bea \lb{dZ-1}
\frac{d\phi}{dA} &=& X, \\
\lb{DX-1}
\frac{dX}{dA} &=& - \frac{1}{(A-A_h)} \left(X+\frac{a^2}{2} \frac{V_{\phi}(\phi)}{V(\phi)} \right), \,
\eea
where \eqref{CZa-in} is used. Eqs. \eqref{dZ-1}-\eqref{DX-1} are easily reduced to a single second order equation
\be \lb{one-eq-XF-1}
(A-A_{h})\frac{d^2 \phi}{dA^2}+\frac{d\phi}{dA}+\frac{a^2}{2} \frac{V_{\phi}(\phi)}{V(\phi)}=0.
\ee
Next, we want to consider a simplification of this equation. For this we expand the last term in a Taylor series up to the first order near one of the extrema of the potential $\phi_{h}=\phi_{1,2,3}$. As a result we obtain
\be \lb{one-eq-XF}
(A-A_{h})\frac{d^2 \phi}{dA^2}+\frac{d\phi}{dA}+\frac{a^2}{2} \, \frac{V_{\phi \phi}(\phi_h)}{V(\phi_h)}(\phi-\phi_h)=0 \,, 
\ee
where we used that $\phi_{h}$ is also a critical point of the potential, i.e.  $V_{\phi}(\phi_h)=0$.

We are looking for solutions to 
the latter equation with the following initial conditions
\be \lb{incond}
\phi(\varepsilon) = \phi_h + \delta \,, \quad \phi'(\varepsilon) = \upsilon \,,
\ee
where $\varepsilon,\delta,\upsilon$ are small, $\varepsilon>0$ and $\delta,\upsilon$ may have an arbitrary sign.   The choice of  the initial conditions (\ref{incond}) means that solutions of our interest are those near the critical points
 $\bar{p}_{1,2,3}$ in the cylinder, see Fig. \ref{Fig:cylinder-case-one}.
It worth to be noted that we assume $\varepsilon^2=\delta^2=0$, but $\upsilon^2$ may not be zero\footnote{In fact, we have already significantly used the first two relations in the derivation of eq. \eqref{one-eq-XF}}.

Solving eq.\eqref{one-eq-XF} we also must take into account the sign of the coefficient in front of $\phi(A)$, i.e. 
\be\label{Deltahi}
\Delta^{(h)}_{i}:=\displaystyle{\frac{a^2}{2}\frac{V_{\phi \phi}(\phi_h)}{V(\phi_h)}},
\ee
with $i=1,2,3$. It is interesting to note that the components of the eigenvalues of the critical points \eqref{c-p-t}  have the same structure \eqref{conVV} as (\ref{Deltahi}). At the same time one can recognise in (\ref{Deltahi}) a combination of the conformal dimension (\ref{confDim}) and the radius of an AdS spacetime $\ell$. Assuming that $\frac{1}{2}<a^2<1$, as  a more general case, we get the conditions 
\be \lb{sign-conf-n}
\Delta^{(h)}_{1} =-2a^2(a^2-1) >0 \,, \quad 
\Delta^{(h)}_{2,3} = 4(a^2-1) < 0.
\ee
The relation for $\Delta^{(h)}_{1}$ (\ref{sign-conf-n}) holds for $0<a^2\leq \frac{1}{2}$.

Let us first consider the case $\phi_h=\phi_1$, i.e. the value of the scalar field on the horizon is the same as at the extremum of the potential related with CFT$_{1}$ UV. Taking into account \eqref{sign-conf-n} the general solution to the scalar field equation \eqref{one-eq-XF} can be written in the form
\be \lb{gen-bes-sol}
\phi(A) = \phi_1+ {\mathrm c_1} J_0\left(2\sqrt{ \Delta^{(h)}_{1} (A-A_h)} \right)+{\mathrm c_2} Y_0\left(2\sqrt{ \Delta^{(h)}_{1} (A-A_h)} \right) \,,
\ee
where $J_{0}$ and $Y_{0}$ are the Bessel functions of the first and second kinds, correspondingly.
Expanding in series \eqref{gen-bes-sol} by $A$ near $A_h$ up to the first order we obtain
\be \lb{Exp-Bes}
\phi (\mathsf{A}) = \phi_1+(1-\mathsf{A}) {\mathrm c_1}+\frac{{\mathrm c_2}}{\pi} \Bigl (2 \gamma + \ln \mathsf{A} - 
    \mathsf{A}  (2 (\gamma-1) + \ln \mathsf{A} ) \Bigr ) \,, 
\ee
where $\gamma$ is Euler–Mascheroni constant and for convenience we introduce $\mathsf{A} := \Delta^{(h)}_{1} (A-A_h)$.
Then the constants ${\mathrm c_{1}}$ and ${\mathrm c_{2}}$  are 
\bea \lb{const-c1}
{\mathrm c_1}& =& \delta+\Bigl( \Delta^{(h)}_{1} \delta (1- \ln (\Delta^{(h)}_{1} \varepsilon) - 2 \gamma) - \upsilon(\ln(\Delta^{(h)}_{1} \varepsilon) + 2\gamma) \Bigr) \varepsilon \,, \\
\,\nonumber\\
\lb{const-c2}
 {\mathrm c_2} &=& \pi \varepsilon (\Delta^{(h)}_{1} \delta + \upsilon). 
\eea


Now we proceed with the reconstruction of the metric functions $A(w)$ and $g(w)$. By virtue of eqs. \eqref{XvarDef} and \eqref{forY-nn}  the function $f(A)$ has the form
\be\label{blackfacts}
f(A)\simeq e^{\mathrm{c_{g}}} (A-A_h),
\ee
where $\mathrm{c_g}$ is a constant of integration. Note, 
that (\ref{blackfacts}) is the same as for the reconstructed AdS black hole solution from the reduced dynamical system with the constant scalar field, see  Appendix \ref{app:AppB2}. 

To find $A$ we subtract eq.~\eqref{eom3} from \eqref{eom1},  that brings us  to the following equation
\be\label{mastEqA-1}
\ddot{A} + \frac{\dot{\phi}^2}{a^2} = 0.
\ee
Using the solution \eqref{Exp-Bes} with \eqref{const-c1}-\eqref{const-c2} and assuming $\varepsilon \sim 0$,
 we obtain the following equation for the scalar factor $A$
\be \lb{mastEqA-2s}
\ddot{A} = 0.
\ee
From the latter we find that the solution for $A$ near the horizon reads
 \be\label{scalsimple}
 A \simeq \sqrt{-\frac{V(\phi_{h})}{2}}w,
 \ee
 where the constants of integration are found as in  Appendix \ref{app:AppB2} and $\phi_{h} =\phi_{1}$.
 Owning to (\ref{blackfacts}),(\ref{scalsimple}) and choosing $\mathrm{c_{g}}=\ln 2$ (see  Appendix \ref{app:AppB2}), the near-horizon  metric takes the form
\be
ds^2 \simeq e^{2\sqrt{-\frac{V(\phi_{h})}{2}}w}\left(-\sqrt{-2 V(\phi_h) } (w-w_h)dt^{2} + dx^2\right) +\frac{dw^2}{\sqrt{-2 V(\phi_h) } (w-w_h)}.
\ee
 

Note, that the assumption $\varepsilon\sim 0$  leads to the special solution with $\mathrm{c_{2}}=0$  (\ref{const-c2}) and
the second initial condition (\ref{incond}) turns to be 
\be
\phi'(0)=-\delta \Delta^{(h)}_{1},
\ee
where $\Delta^{(h)}_{1}$ defined by \eqref{sign-conf-n}.
This yields that the solution to \eqref{Exp-Bes} for the scalar field is a linear function
\be\label{sfsimpSol}
\phi \simeq  \left(1-\Delta^{(h)}_{1} \sqrt{-\frac{V(\phi_{h})}{2}}(w-w_{h})\right)\delta.
\ee

We see that the near-horizon form of the scale factor $A$ (\ref{scalsimple}) and the blackening function $f$ (\ref{blackfacts})  are the same as those for a near-horizon AdS black hole solution found in Appendix \ref{app:AppB2}. The scalar field (\ref{sfsimpSol}) vanishes as we send $\delta$ to zero $\delta\to 0$. 

In the same way we are able to find similar representation for the metric functions and scalar field for the case when the value of the scalar field on the horizon is as at the other critical points of the potential, $\phi_{h}=\phi_{2,3}$.

\subsection{Generalization for an arbitrary $\phi_{h}$}

In this subsection we obtain a near-horizon dilatonic black hole solution for our model when the value of the scalar field at the horizon $\phi_h$ is not necessarily the value at the critical point of  $V(\phi)$.

The Taylor expansion of the last term in eq.\eqref{one-eq-XF-1} near an arbitrary $\phi_{h}$ yields 
\bea
\lb{ppp-ex}
\frac{a^2}{2}\frac{V_{\phi}(\phi)}{V(\phi)}\Big|_{\phi_{h}}&\cong & \frac{a^2}{2}\left(\frac{V_{\phi}(\phi_{h})}{V(\phi_{h})} + \left(\frac{V_{\phi\phi}(\phi_{h})}{V(\phi_{h})}-\frac{V_{\phi}(\phi_{h})^2}{V(\phi_{h})^2}\right) (\phi -\phi_{h})\right),\\
&= & \mathsf{\Lambda}^{(h)}+\left(\Delta^{(h)}-\frac{2}{a^2}(\mathsf{\Lambda}^{(h)})^2\right)(\phi-\phi_{h}),\\
&=&\mathsf{\Lambda}^{(h)}+\mathsf{K}^{(h)}(\phi-\phi_{h}),
\eea
where we denote 
\be \lb{SlDef}
\mathsf{\Lambda}^{(h)}=\displaystyle{\frac{a^2}{2}\frac{V_{\phi}(\phi_h)}{V(\phi_h)}},\quad  \Delta^{(h)} =\frac{a^2}{2}\frac{V_{\phi\phi}(\phi_{h})}{V(\phi_{h})},\quad\mathsf{K}^{(h)}=\left(\Delta^{(h)}-\frac{2}{a^2}\left(\mathsf{\Lambda}^{(h)}\right)^2\right).
\ee

Note, that for $a^2\leq \frac{1}{2}$ the quantity $\mathsf{K}^{(h)}$ is  always non-negative, while for $\frac{1}{2}<a^2<1$ we have to trace the sign of  $\mathsf{K}^{(h)}$ with respect to the position of $\phi_{h}$ and also take into account zeros of the scalar potential. Using \eqref{ppp-ex}-\eqref{SlDef} in \eqref{one-eq-XF-1} we get
\be \lb{ppp-ex-eq}
(A-A_{h})\frac{d^2 \phi}{dA^2}+\frac{d\phi}{dA}+\mathsf{K}^{(h)}(\phi-\phi_{h}) +\mathsf{\Lambda}^{(h)}=0.
\ee
The solution to eq. \eqref{ppp-ex-eq} can be  represented as follows
\bea\label{solDilStar}
\phi(A) =
\begin{cases}
\mathrm{c_1} J_0\left(2 \sqrt{\mathsf{K}^{(h)}(A-A_{h}) }\right)+ \mathrm{c_2} Y_0\left(2 \sqrt{\mathsf{K}^{(h)}(A-A_{h})}\right)+\phi_{h} -\frac{\mathsf{\Lambda}^{(h)} }{\mathsf{K}^{(h)}},&\text{for}\quad \mathsf{K}^{(h)}>0, \\
\,\\
\mathrm{c_1} I_0\left(2 \sqrt{|\mathsf{K}^{(h)}(A-A_{h})|}\right)+ \mathrm{c_2} K_0\left(2 \sqrt{|\mathsf{K}^{(h)}(A-A_{h})| }\right)+\phi_{h} -\frac{\mathsf{\Lambda}^{(h)} }{\mathsf{K}^{(h)}},&\text{for}\quad \mathsf{K}^{(h)}<0\,,
\end{cases}
\eea
where $I_{0}$ and $K_{0}$ are modified Bessel functions of the first and second kinds, correspondingly. 
The Taylor expansion (\ref{solDilStar}) near $A_{h}$  up to the first order regardless of the sign of $\mathsf{K}^{(h)}$ can be represented in the form 
\bea \lb{gen-phi}
\phi(\bar{\mathsf{A}}) = \phi_{h}-\frac{ \mathsf{\Lambda}^{(h)}}{ \mathsf{K}^{(h)}}+ \mathrm{\tilde{c}_1}(1-\bar{\mathsf{A}})+ \mathrm{\tilde{c}_2}\left(2 \gamma+\ln|\bar{\mathsf{A}}| - \bar{\mathsf{A}}(2(\gamma - 1)+\ln|\bar{\mathsf{A}}|)\right) \,, 
\eea
where $\bar{\mathsf{A}}=\mathsf{K}^{(h)}(A-A_{h})$ and
 the constants of integration $\mathrm{\tilde{c}_{1}}, \mathrm{\tilde{c}_{2}}$ are chosen in the form
\be \lb{genconst}
\mathrm{\tilde{c}_1} = \frac{\mathsf{\Lambda}^{(h)} }{\mathsf{K}^{(h)}}+ \varepsilon\left( 
\mathsf{\Lambda}^{(h)}
-2 \gamma( \mathsf{\Lambda}^{(h)} +  \upsilon)  -  (\mathsf{\Lambda}^{(h)} +\upsilon) \ln (|\mathsf{K}^{(h)} \varepsilon |)\right), \quad \mathrm{\tilde{c}_2} = \varepsilon  (\mathsf{\Lambda}^{(h)} +\upsilon).
\ee

To find the scale factor $A$  we use the equation (\ref{mastEqA-1}), which is rewritten as follows
\be\label{mastEqA-2}
\ddot{A} + \dot{A}^2\left(\frac{d\phi(A)}{dA}\right)^2 = 0.
\ee
Plugging only the leading term of the quantity $\displaystyle{\Bigl(\frac{d\phi}{d A}\Bigr)^2}$ with $\varepsilon =0$ into eq.(\ref{mastEqA-2}) , we find that (\ref{mastEqA-2}) reduces to
\be
\ddot{A} + \frac{\mathsf{\Lambda}^{(h)}}{a^2}\dot{A}^2 = 0.
\ee
Then, the  solution for $A$ reads
\be \lb{AwithC}
A(w)= \frac{a^2}{(\mathsf{\Lambda}^{(h)})^2} \ln \left(\frac{(\mathsf{\Lambda}^{(h)})^2}{a^2} w+\mathrm{C_{1}}\right)+\mathrm{C_{2}},
\ee
where $\mathrm{C_1}$ and $\mathrm{C_2}$ are  constants of integration, which will be determined below from fairly simple considerations.
The solution for $A$ (\ref{AwithC}) should reproduce the near-horizon AdS black hole solution \eqref{ck-sol} if $\phi_{h}$ is the critical point of the potential. 
Remembering that for such points  of the potential $\mathsf{\Lambda}^{(h)} = 0$, since $V'(\phi_{h})=0$, we expand in series the scale factor \eqref{AwithC} assuming that the parameter $\mathsf{\Lambda}^{(h)}$ is small, as a result we obtain
\be \lb{sefA}
A(w) \simeq \mathrm{C_{2}} + \frac{\ln(\mathrm{C_{1}})a^2}{(\mathsf{\Lambda}^{(h)})^2}+\frac{w}{\mathrm{C_{1}}} + o\left((\mathsf{\Lambda}^{(h)})^2\right).
\ee
Thus, choosing the integration constants $\mathrm{C_{1}}$ and $\mathrm{C_{2}}$ in the following form
\be \lb{fcfa}
\mathrm{C_{2}}= -\frac{\ln(\mathrm{C_{1}})a^2}{(\mathsf{\Lambda}^{(h)})^2} \,, \quad \mathrm{C_{1}} = \sqrt{-\frac{2}{V(\phi_{h})}},
\ee
and substituting them into eq.\eqref{sefA}, we automatically recover the  scale factor of the near-horizon AdS solution \eqref{ck-sol}.
 From (\ref{AwithC}) with  $\mathrm{C}_{2}$ (\ref{fcfa}) we find
\be \lb{fsolA}
A(w) = \frac{a^2}{(\mathsf{\Lambda}^{(h)})^2} \ln \left(\frac{\frac{(\mathsf{\Lambda}^{(h)})^2}{a^2} w+\mathrm{c_{A}}}{\mathrm{c_{A}}}\right) \,,
\ee
where we redefine $\mathrm{C_1}$  as $\mathrm{c_A}$ for convenience.

Now we are able to restore the blackening function $f$ near the horizon. By virtue of \eqref{forY-nn} and the definition of $Y$ (\ref{XvarDef}),   the function $g(A)$ reads
\be \lb{forGA}
g(A) = \mathrm{c_g}+\ln(A-A_h).
\ee
Note, that $g$ given by (\ref{forGA})  matches with $g$ for the exact AdS black hole \eqref{ggA-s} in terms of $A$, however, the constants $\mathrm{c_g}$ in (\ref{forGA}) and \eqref{ggA-s} are different.
  We find $\mathrm{c_g}$ in (\ref{forGA}) as it was done for the AdS black hole in Appendix \ref{app:AppB2}. Using the definition of $Y$ (\ref{XvarDef})
we find that eq.\eqref{eom1} near horizon turns to be
\be \lb{eqNhor}
\dot{A}^2(w) \, Y(A(w)) + e^{-g(A(w))}V(A(w)) = 0,
\ee
where $\phi$ and $g$ are represented as functions of $A$ and only the leading terms are left. Similarly, we come from the full $3d$ dynamical system \eqref{dz/da}-\eqref{dy/da} to eqs.\eqref{Aeq-dz-dx}-\eqref{Aeq-dz-dx-1} with $Y$ tending to infinity. Substituting into eq. \eqref{eqNhor} $Y$ and $g$ given by \eqref{forY-nn} and \eqref{forGA}, respectively,  we come to
\be \lb{eqNhor-1}
\dot{A}^2(w) \, \frac{1}{A(w)-A_h} + V(A(w))e^{-\mathrm{c_g}} \frac{1}{A(w)-A_h} = 0.
\ee
Taking into account the scale factor (\ref{fsolA}) the latter equation is brought to 
\be \lb{eqNhor-2}
\frac{1}{\left(\frac{(\mathsf{\Lambda}^{(h)})^2}{a^2} w+\mathrm{c_{A}}\right)^2} + V(A(w))e^{-\mathrm{c_g}} = 0 \,. 
\ee
Since we are looking for a near-horizon solution, only two terms from eq.\eqref{eom1} dominate in eq.\eqref{eqNhor-1}). Expanding in series eq.\eqref{eqNhor-2} near $w_h$,  we get 
\bea \lb{eqNhor-2m}
\frac{1}{\left(\frac{(\mathsf{\Lambda}^{(h)})^2}{a^2} w_h+\mathrm{c_{A}}\right)^2} \left (1-\frac{2}{a^2}\frac{(\mathsf{\Lambda}^{(h)})^2}{\left(\frac{(\mathsf{\Lambda}^{(h)})^2}{a^2} w+\mathrm{c_{A}}\right)^2} (w-w_h) \dots \right) = \nonumber\\[7pt]
e^{-\mathrm{c_g}} |V(\phi_h)| \left (1-\frac{2}{a^2}\frac{(\mathsf{\Lambda}^{(h)})^2}{\left(\frac{(\mathsf{\Lambda}^{(h)})^2}{a^2} w+\mathrm{c_{A}}\right)^2} (w-w_h) \dots \right) \,,
\eea
where 
we use the definition of the parameters \eqref{SlDef} and the solution for the scalar field \eqref{gen-phi}-\eqref{genconst}. Finally, we find for $\mathrm{c_g}$
\be \lb{CGE}
\mathrm{c_g} = \ln \left (|V(\phi_h)|\left(\mathrm{c_A}+\frac{(\mathsf{\Lambda}^{(h)})^2}{a^2} w_h\right)^2 \right ) \,
\ee
or, taking into account (\ref{SlDef}), (\ref{fcfa}), we  can  represent (\ref{CGE}) as follows
\be
\mathrm{c_g} =\ln\left(2+ \frac{a^2}{\sqrt{2}} \frac{|V_{\phi}(\phi_{h})|^{2}}{|V(\phi_h)|^{3/2}}w_h+ \frac{a^{4}}{16}\frac{|V_{\phi}(\phi_{h})|^{4}}{|V(\phi_{h})|^{3}}w_{h}^{2}\right),
\ee
from which one can see that choosing $\phi_{h}$ as $\phi_{h}=\phi_{1,2,3}$, the constant $\mathrm{c_g}$  turns to be $\mathrm{c_{g}}=\ln2$ as for the near-horizon metric of the AdS black hole  \eqref{forAlog}.

Therefore, the blackening function $f$ is given by
\be\label{fsolmain}
f =  \frac{a^2 \mathrm{c_{g}}}{(\mathsf{\Lambda}^{(h)})^2} \ln \left(\frac{\frac{(\mathsf{\Lambda}^{(h)})^2}{a^2} w+\mathrm{c_{A}}}{\frac{(\mathsf{\Lambda}^{(h)})^2}{a^2} w_{h}+\mathrm{c_{A}}}\right) \,, \ee
where $\mathrm{c_{g}}$ is defined by (\ref{CGE}).
The black hole solution near the horizon is given by
\be\label{solmetrF}
ds^2 \simeq  \left(\frac{\frac{(\mathsf{\Lambda}^{(h)})^2}{a^2} w+\mathrm{c_{A}}}{\mathrm{c_{A}}}\right)^{\frac{2a^2}{(\mathsf{\Lambda}^{(h)})^2}}\left(-f dt^2 +dx^2\right)+\frac{dw^2}{f},
\ee
with $f$ given by  (\ref{fsolmain}), and the scalar field  by (\ref{gen-phi}) with (\ref{genconst}).
It is interesting to note that the near-horizon solutions (\ref{solmetrF}),(\ref{gen-phi}), which we have constructed, have scaling geometries similar to those considered in \cite{Goldstein:2009cv}.

 \setcounter{equation}{0}
\section{Improved near-horizon solutions for the scalar field}\label{sec:solPhiImp}

In this section we will present  a sketch how to find analytic solutions for asymptotically AdS black holes using  additional conditions, which allow to construct a solution for the scalar field $\phi(A)$,  defined for both near-horizon and near the boundary regions unlike the solution \eqref{solDilStar}. 

Imposing the constraint $X^2 \sim 0$ for the original equations \eqref{dz/da}-\eqref{dy/da} we are brought to
 the following system of equations
\bea\label{adz/da}
&&\frac{dz}{dA}=z(z-1)X,\\ \label{adx/da}
&&\frac{dX}{dA}=-\left(Y+2\right)\left(X+\mathcal{C}_{(z,a)} \right),\\ \label{ady/da}
&&\frac{dY}{dA}=-Y\left(Y+2\right) \,.
\eea

The system \eqref{adz/da}-\eqref{ady/da} can also be projected into the unit cylinder, so the solutions to it will almost coincide with the flows\footnote{Now we are mainly interested in flows ending at $p_1$.} described in Figs. \ref{Fig: CrPCzPos2-025} and \ref{Fig:cylinder-case-one}.
It is interesting to note that for the case $0<a^2<1/2$ the solutions to eqs. \eqref{adz/da}-\eqref{ady/da} in the unit cylinder, which start with some $\phi_h$ and end at the critical point $p_{1}$ have a greater deviation from the trajectories in Fig. \ref{Fig: CrPCzPos2-025}, which also end at $p_1$, the further from $z=1/2$ (i.e. the extremum of $V(\phi)$ at the origin) the initial condition is chosen. This happens due to the fact that the corresponding flows deviate more from $x=0$. 

Let us now consider the solution of the system \eqref{adz/da}-\eqref{ady/da} assuming that $Y(0)$ tends to $\infty$. The equation \eqref{ady/da} is independent and has the following solution (see \eqref{dy/da-r-s}, also we take $A_h=0$ for simplicity)  
\be \lb{forAY}
Y(A) = \frac{2}{e^{2 A}-1} \,.
\ee
Substituting the solution \eqref{forAY} into eq.\eqref{adx/da} we come to
\bea \lb{adZ-1}
\frac{d\phi}{dA} &=& X, \\
\lb{aDX-1}
\frac{dX}{dA} &=& -\left(1+\coth A \right) \left(X+\frac{a^2}{2} \frac{V_{\phi}(\phi)}{V(\phi)} \right) \,,
\eea
where as in the previous section it is convenient to return from $z$ to the original variable $\phi$ \eqref{Zvar}.
Eqs. \eqref{adZ-1}-\eqref{aDX-1} are reduced to a single second-order differential equation
\be \lb{a-one-eq-XF-1}
\frac{d^2 \phi}{dA^2}+\left(1+\coth A \right) \left(\frac{d\phi}{dA}+\frac{a^2}{2} \frac{V_{\phi}(\phi)}{V(\phi)}\right ) =0 \,.
\ee
It worth to be noted that the equation \eqref{one-eq-XF-1} for the scalar field from the previous section is just a series expansion of eq.\eqref{a-one-eq-XF-1} by small $A$ up to first order.
Next, expanding  eq.\eqref{a-one-eq-XF-1} near $\phi_h$ and doing some algebra, we get the equation
\be \lb{th-eq-XF}
\tanh A \,  \frac{d^2 \Phi}{dA^2}+\left(\tanh A+ 1\right) \left(\frac{d\Phi}{dA}+\mathsf{K}^{(h)}\Phi \right ) =0 \,,
\ee
where we inroduce the following quantity
\be \lb{PHIde}
\Phi:=\phi-\phi_{h}+\frac{\mathsf{\Lambda}^{(h)}}{\mathsf{K}^{(h)}},
\ee
with $\mathsf{\Lambda}^{(h)}$ and $\mathsf{K}^{(h)}$ defined by \eqref{SlDef}. 
Eq.(\ref{th-eq-XF}) with  the transformation 
\be \lb{th-eq-XF-t}
r=e^{2A}
\ee
can be yielded the form
\be \lb{HG}
r(1-r)\, \frac{d^2 \Phi(r)}{dr^2}+(1-2r)\,\frac{d\Phi(r)}{dr}-\frac{\mathsf{K}^{(h)}}{2} \Phi(r) =0.
\ee
The latter is a hypergeometric equation whose regular singular points $r=1$ and $r=\infty$ correspond to the near-horizon and boundary regions, respectively.\footnote{Recall that the scale factor $A$ tending to $0$ is related with a solution near the horizon, while $A \to \infty$ is associated with a solution near the boundary.}.
One of the fundamental solutions of eq. (\ref{HG}) near $r=1$ is represented in the form 
\be \lb{2f1}
\Phi(r)={}_2 F_1 (a_h,1-a_h,1,1-r) \,,
\ee
where the parameter $a_{h}$ is defined as follows
\be\label{hypgpar}
a_h=\frac{1}{2}\left(1-\sqrt{1-2\mathsf{K}^{(h)}}\right).
\ee
One can check that the solution for the scalar field (\ref{PHIde}) with (\ref{2f1}) is related with the branch of solutions \eqref{solDilStar} given by Bessel functions of the first 
kind. This follows from the Taylor expansions
\bea \lb{BEf}
J_{0}\left(2\sqrt{\mathsf{K}^{(h)} A }\right) &=& 1 - \mathsf{K}^{(h)} A + \frac{(\mathsf{K}^{(h)} A)^2}{4} - \frac{(\mathsf{K}^{(h)} A)^3}{36}+\dots \\[7pt]
\nonumber
{}_2 F_1 (a_h,1-a_h,1,1-e^{2A}) &=& 1 - \mathsf{K}^{(h)} A + \frac{(\mathsf{K}^{(h)} A)^2}{4} - \frac{(\mathsf{K}^{(h)} A)^3}{36} \, \left(\frac{\mathsf{K}^{(h)}-2}{\mathsf{K}^{(h)}}\right)  +\dots .
\eea


We should make some remarks regarding the limits of applicability of the equation \eqref{th-eq-XF}. This equation corresponds to the system \eqref{adz/da}-\eqref{ady/da}, for which the function $\mathcal{C}_{(z,a)}$ is expanded in series by $z$ near $z_h$ up to the first order. The flows associated with the simplified system \eqref{adz/da}-\eqref{ady/da} in the unit cylinder under the initial conditions of type \eqref{in-con-nnx}  almost perfectly coincide with the flows in Fig. \ref{Fig:cylinder-case-one} but only if $z_h \sim z_1$; in other cases, there will be a good match as long as $y\neq 0$.

It should be noted that the equation (\ref{HG}) is very similar to the equation for the holographic two-point function for the holographic model (\ref{act}) from \cite{Deger:2002hv} (up to the evolution parameter).

\section{Conclusions}\label{sec:concl}

In this work we have consider holographic RG flows at finite temperature in $D=3$ $\mathcal{N}=(2,0)$ gauged supergravity coupled to a sigma model. Depending on the parameter $a^2$ related to the curvature of the target space, the scalar potential of the holographic model can have  either a single ($0<a^2\leq \frac{1}{2}$) or three extrema ($\frac{1}{2}<a^2<1$) giving rise to AdS solutions, which are dual to CFTs. 

We have represented the equations of motion of the holographic model as the autonomous dynamical system.  Certain trajectories of the system can be interpreted as  RG flows, while equilibria of the system correspond to  fixed points of RG flow at zero temperature. Because of the black hole ansatz for the metric the dynamical system has infinite points corresponding to the near-horizon regions. By virtue of the Poincar\'e transformations we have projected the system into the unit cylinder, so the near-horizon regions mapped onto the critical line on the boundary of the cylinder. 

We have numerically solved the equations of the system in the cylinder and plotted the global phase portraits. The flows starting from the near-horizon regions to the  spacetime boundaries with various $\phi_{h}$ (similarly to \cite{Gursoy:2018umf,Bea:2018whf}) have been found numerically. 
Using the autonomous dynamical system helps to understand the behavior of the trajectories based on the stablity analysis of fixed points.
For the potential with a single extremum asymptotically AdS black hole solutions with any value of $\phi_h$ pass close to the exact RG flow, which is a separatrix, and end at the AdS fixed point, see Fig.~\ref{Fig: CrPCzPos2-025}. 
For the potential with a three extrema, the separatrices are holographic RG flows between AdS fixed points, corresponding to different extrema of the scalar potential, which were constructed numerically in \cite{Deger:2002hv, NE-RG}, see Fig.~\ref{Fig: SyclineAdS2}.
We have observed that black hole solutions, for which the value of $\phi_{h}$ is close to one of the additional extrema of the potential $\phi_{2,3}$, can end either at the AdS fixed point related with these extrema ($\phi_{h}=\phi_{2,3}$) or at the AdS fixed point associated with the maximum of the potential ($\phi_h\neq  \phi_{2,3}$), see Fig.~\ref{Fig:cylinder-case-one}. Another type of solutions, which we have seen setting $\phi_{h}$ near the additional extrema $\phi_{2,3}$, are naked singularities.

We have found an analytical description of asymptotically AdS black holes for the near-horizon region. The constructed near-horizon geometries are scaling and defined by the value of the scalar field at the horizon $\phi_{h}$. For $\phi_h$ coinciding with one of the extrema of the scalar potential the solution reproduces the metric of the AdS black hole near the horizon. 

To extend the analytic black hole solutions to the boundary we have used additional constraints for the dynamical equations. This has led us to a hypergeometric equation for the scalar field on the scale factor. One of the regular singularities of this equation corresponds to the near-horizon region.

Despite we have considered a very special three-dimensional holographic model with the certain scalar potential, the method of construction and analysis of RG flows at finite temperature can be generalized to an arbitrary dimension similarly to  \cite{Kiritsis:2016kog,Gursoy:2018umf} as long as we consider black hole generalizations of Poincar\'e invariant domain wall solutions. 

A natural continuation of this work is to find the full solutions of asymptotically AdS black holes from the horizon to the boundary, next, using them, to explore in details thermodynamical properties of RG flows. 
It will be interesting to construct and analyze a finite temperature generalization of an exotic RG flow for our model, which was found in \cite{NE-RG}. \\

\section*{Acknowledgements}
We are grateful to Irina Ya. Aref'eva, Maria A. Skugoreva  and Nikolay A. Tyurin for helpful discussions. We thank Eric Gourgoulhon for collaboration at the initial stages of this work. 
A.G. is grateful to the Paris Observatory, LUTh for warm hospitality, where a part of this work was done. The work of A.G. presented in Sections 2, 4 was supported by Russian Science Foundation grant RSCF-22-72-10122. The work of M.P. was supported by the JINR AYSS Foundation, Project No 24-301-03.

\newpage

\appendix
\setcounter{equation}{0} \renewcommand{\theequation}{A.\arabic{equation}}

\section{Exact holographic RG flow}\label{app:AppA}
\setcounter{equation}{0}\renewcommand{\theequation}{A.\arabic{equation}}
For the model (\ref{act}) the exact domain wall solution, which preserve half of supersymmetries exists and was obtained in the work \cite{Deger:2002hv}. 
The scale factor of this solution reads
\be\label{degerSF}
A_{\rm susy} = \frac{1}{4a^2}\ln(e^{4a^2w}-1),
\ee
the scalar field is given by
\be\label{degerScF}
\phi_{\rm susy} =  \frac{1}{2}\ln\left(1+\frac{e^{-2a^2w}}{1-e^{-2a^2w}}\right),
\ee
where the radial coordinate $w$ runs from $0$ to $\infty$.\\

The metric of the domain wall solution with (\ref{degerSF}) can be represented as follows
\be\label{metricDeg}
ds^2_{\rm susy} = \left(e^{4a^2w} -1 \right)^{\frac{1}{2a^2}}\left(-dt^2+dx^2\right) +dw^2.
\ee
The domain wall solution (\ref{metricDeg}) with (\ref{degerScF}) has an AdS asymptotics with $w\to \infty$, while if one send  $w$ to zero, the  solution  has a singular geometry. For $a^2\leq\frac{1}{2}$  this singular behaviour turns to be acceptable since the solution satisfies the Gubser's bound.

\section{AdS black hole solutions}\label{app:AppBM}
\setcounter{equation}{0}\renewcommand{\theequation}{B.\arabic{equation}}

\subsection{Exact solutions} \label{app:AppB}
Before analyzing the general case, we show that we are able to reconstruct an AdS black hole solution (\ref{ck-sol}) from a solution of the autonomous system (\ref{dz/da})-(\ref{dy/da}). Setting the scalar field to be constant $\phi = \phi_{h}$ (it can be $\phi_1=0$ or $\phi_{2,3}$ from (\ref{crP-v})) and assuming that the scale factor $A \neq const$,  it follows  from 
(\ref{XvarDef})-(\ref{Zvar}) that $X=0$ and $Z=const$. 

Then the system (\ref{dz/da})-(\ref{dy/da}) 
is reduced to the one equation on $Y$
\be \lb{dy/da-r}
\frac{dY}{dA} = -Y \biggl(Y+2 \biggl) \,. 
\ee
To solve eq.(\ref{dy/da-r}) we impose the initial condition 
\be
Y(A_h) = \delta,
\ee
with  $A_h=A(w_h)$, $\delta>0$, that leads to the following solution
\be \lb{dy/da-r-s}
Y(A) =  \frac{2 \delta}{ (\delta + 2) e^{2(A-A_h)}-\delta} \,.
\ee
Considering the limit $\delta \to \infty$ in (\ref{dy/da-r-s}), we obtain 
\be \lb{dy/da-r-sl}
Y(A) = \frac{2}{e^{2 (A-A_h)}-1}\,. 
\ee

Substituting (\ref{dy/da-r-sl}) into the relation for $Y$
 (\ref{XvarDef}) and 
integrating the resulting expression, we find the $log$ of the blackening function $f$, i.e. $g(A)$:
\be \lb{dg/da-s}
g = {\rm c_g} + \ln \left(1 -e^{-2(A-A_h)}\right),
\ee
where ${\rm c_g}$ is a constant of integration.

One can find a solution to $A$ in terms of $w$ integrating the equation
\be
\ddot{A}=0,
\ee
that immediately gives
\be\label{exactAsol}
A = {\rm c_{A}}w + {\rm c_{w}}.
\ee
The latter integration constant ${\rm c_{w}}$ in (\ref{exactAsol}) can be set to $0$ due to  the equations of motion are invariant under shift of the radial coordinate $w$. As for ${\rm c_{A}}$ we can derive it plugging back  (\ref{dg/da-s}),(\ref{exactAsol}) into any of the Einstein equations (\ref{eom1})-(\ref{eom3}), so
\be\label{constcaExp}
{\rm c_{A}} =e^{-\rm c_{g}/2} \sqrt{-\frac{V(\phi_{h})}{2}}.
\ee
The constant $\rm c_g$ can be find  taking into account that the blackening function $f$ on the boundary should tend to $1$. Thus, ${\rm c_{g}}$ is fixed as
\be
{\rm c_{g}}=0.
\ee
Note, that the condition that  $f$ goes to $0$ on the horizon is also satisfied with \eqref{dg/da-s}. 
Finally, the metric functions look like
\be
A = \sqrt{-\frac{V(\phi_{h})}{2}} w, \quad f = 1 -e^{-\sqrt{-2V(\phi_h)}(w-w_h)},
\ee
that is in agreement with \eqref{ck-sol}.

We also show how to find a near-horizon asymptotic solution  using the  reduced dynamical system below. For certain choice of integration constants we see matching between the exact and asymptotic solutions.

\subsection{Near-horizon solutions} \label{app:AppB2}


Next, we will demonstrate, using as an example eq. \eqref{dy/da-r}, how to recover gravitational solutions that matches to solutions of a corresponding dynamical system near its infinite critical points.

For (\ref{dy/da-r}) we find two finite equilibrium points 
\be
Y=0, \quad Y=-2.
\ee

To analyze the system at infinity we map (\ref{dy/da-r}) to $1d$ disc $\mathbf{D}$, using transformation
\be\label{PoincareMap2}
 Y = \frac{\tilde{Y}}{\sqrt{1-\tilde{Y}^2}},
\ee

Then the dynamical system (\ref{dy/da-r}) on the disc looks like
\be\label{onedsdisc}
\frac{d\tilde{Y}}{dA} = -\tilde{Y}\left(\tilde{Y}\sqrt{1-\tilde{Y}^2}+2(1-\tilde{Y}^2)\right).
\ee

The critical points  corresponding to (\ref{onedsdisc}) are
\be 
\tilde{Y} = 0,\quad \tilde{Y}=-\frac{2}{\sqrt{5}},\quad  \tilde{Y}= -1, \quad \tilde{Y}=1.
\ee

Two additional points $\tilde{Y}=\pm 1$ correspond to $Y=\pm \infty$. Note that the exact solutions indicated in the previous subsection correspond to the solution connecting the critical point $\tilde{Y}=1$ with the critical point $\tilde{Y}=0$.

To find an asymptotic solution to eq.(\ref{onedsdisc}) near the equilibrium point $\tilde{Y} = 1$, we introduce a new variable $\epsilon_{y}$, so that
\be\label{tildeYeps}
 \tilde{Y} = 1-\epsilon_{y},
\ee
where $0<\epsilon_y \ll 1$. Substituting (\ref{tildeYeps}) into (\ref{onedsdisc}) and removing higher order terms, we get
\be\label{e-e-eq}
\frac{d\epsilon_{y}(A)}{dA} \simeq\sqrt{2\epsilon_{y}(A)},
\ee
where we used that $\epsilon$ is small. 

Solving equation (\ref{e-e-eq}) with the boundary 
condition  $\epsilon_{y}(A_h)=\delta$ and then 
sending $\delta$ to zero  we find
\be \lb{e-e-sol}
\epsilon_{y}(A) \simeq \frac{1}{2} (A-A_h)^2.
\ee
Doing the inverse transformations (\ref{tildeYeps}) and (\ref{PoincareMap2}), we obtain
\be
Y \simeq \frac{1}{A-A_h} -\frac{1}{2} (A-A_h) \,, 
\ee
where $A \gtrsim A_h$. Remembering the definition of $Y$ from the second relation of (\ref{XvarDef}) 
we find the corresponding function $g(A)$ 
\be \lb{ggA-s}
g(A) \simeq {\rm c_g} +\ln (A - A_h)\,,
\ee
where ${\rm c_g}$ is a constant  of integration.
So we have for the blackening function $f$
\be \lb{fa-s}
f(A) \simeq e^{{\rm c_g}}(A - A_h).
\ee
To reconstruct the function $A(w)$, we use the Einstein equations setting the scalar field to be constant.
Subtracting (\ref{eom1}) from (\ref{eom3}) and taking into account  $\phi_h = \phi(w_{h}) $, we obtain the equation for $A$
\be \lb{ddA-e}
\ddot{A} =0.
\ee
Then the scale factor $A$ takes the form as in the previous subsection (\ref{exactAsol}) $A={\rm c_{A}}w$. One can determine the constants ${\rm c_{A}}$, ${\rm c_{g}}$ by substitution of the solutions for $A$ and $g$ into any of the  equations of motion (\ref{eom1})-(\ref{eom3}) such that only the terms of higher order survive. Finally, we have for them
\be
{\rm c_{A}} = e^{-{\rm c_g}/2}\sqrt{-V(\phi_{h})}.
\ee
Note that the relation for  ${\rm c_{A}}$ differs by a factor $\frac{1}{\sqrt{2}}$ comparing to the explicit case \eqref{constcaExp}. Thus, to reach matching the solutions we should set ${\rm c_g}$ as follows
\be \lb{forAlog}
{\rm c_{g}} = \ln 2.
\ee

Thus we obtain the following formula for the blackening function near the horizon
\be \lb{gwaa-s}
f \simeq  \sqrt{-2 V(\phi_h) } (w-w_h) .
\ee
Note, that in this case the solution \eqref{gwaa-s} is just an expansion of the solution \eqref{ck-sol} near the horizon.


\setcounter{equation}{0}

\section{Fixed points and holographic RG flows at $T=0$ }\label{app:AppCM}

\subsection{Stability of the critical points}\label{app:AppC}

Here we discuss stability of the critical points of (\ref{EqsonCyl001})-(\ref{EqsonCyl0013}) in sense of the 3d dynamical  system. We summarize the types  of stability for each of the point in  Table.\ref{table: pointsclass}  

\begin{table}[h]
\begin{center}
\begin{tabular}{|c|c|c|c|c|c|}
\hline & $0<a^2<\frac{1}{2}$ & $a^2=\frac{1}{2}$ & $\frac{1}{2}<a^2<1$ & $a^2 = 1$ & $a^2 > 1$ \\
\hline$p_1$ & \multicolumn{3}{|c|}{ stable node }  & non-hyperbolic 
& saddle \\
\hline$p_{2,3}$ & \multicolumn{2}{|c|}{—} & saddle & \multicolumn{2}{|c|}{—} \\
\hline$p_{4,7}$ & saddle & -- &\multicolumn{3}{|c|}{unstable node} \\
\hline $p_{5,9}, p_{6,8}$ & \multicolumn{5}{|c|}{non-hyperbolic, non-isolated} \\
\hline$q_{1,2}$ & — & saddle & \multicolumn{3}{|c|}{—} \\
\hline $g_{1,3}$ & \multicolumn{5
}{|c|}{saddle*} \\
\hline $g_{2,4}$ & \multicolumn{5
}{|c|}{stable node*}\\
\hline
\end{tabular}
\caption{Classification of critical points of the dynamical system (\ref{EqsonCyl001})-(\ref{EqsonCyl0013}).
}
\label{table: pointsclass}
\end{center}
\end{table}

Here the non-hyperbolic means that one of eigenvalues equals to zero, non-isolated means that this point lies on the critical curve. However, if we fix $y=0$, that is, we consider the invariant manifold $xz$-plane, then non-isolated points become isolated and the classification obtained in the works \cite{Golubtsova:2022hfk, NE-RG} is valid for them. The asterisk means that one of the eigenvalues is not defined; but it turns out to be defined in any arbitrarily small neighborhood of these points. The absolute value of this eigenvalue, of course, depends on the neighborhood, but its sign is always uniquely determined. This makes it possible to identify the nature of the stability of these points (saddle, stable node), which was also verified by numerical construction of flows near these points.

\subsection{Holographic RG flows at zero temperature}\label{app:AppC2}

Let's list possible holographic RG flows for different choices of $a^2$:
\begin{enumerate}
    \item $0<a^2<\frac{1}{2}$
    \begin{enumerate}
        \item  from $p_1$ (CFT$_{1}$ {\bf UV})  at $\phi_{*1}=0$ to $p_4$, with $\phi \to +\infty$, triggered by a relevant operator (Dirichlet b.c.),
         \item  from $p_1$ (CFT$_{1}$ {\bf UV})  at $\phi_{*1}=0$ to $p_4$, with $\phi \to +\infty$, triggered by a relevant operator (Neumann b.c.),
        \item  from CFT$_{1}$  {\bf UV} $p_{1}$ at $\phi_{*1}$ to $p_5$, with $\phi \to +\infty$, driven by a non-zero vacuum expectation value of the scalar operator (Dirichlet b.c.),
         \item  from CFT$_{1}$  {\bf UV} $p_{1}$ at $\phi_{*1}$ to $p_5$, with $\phi \to +\infty$, driven by a non-zero vacuum expectation value of the scalar operator (Neumann b.c.);
    \end{enumerate}
    \item $a^2=\frac{1}{2}$
\begin{enumerate}
\item from $p_1$ (CFT$_{1}$ {\bf UV}) at $\phi_{*1}=0$ to $p_5$, with $\phi \to +\infty$, triggered by a non-zero vacuum expectation value of the scalar operator (Dirichlet b.c.);
\item \ from $p_1$ (CFT$_{1}$ {\bf UV}) at $\phi_{*1}=0$ to $p_4$, with $\phi \to +\infty$, triggered by a source (Neumann b.c.);
\item from $p_1$ (CFT$_{1}$ {\bf UV}) at $\phi_{*1}=0$ to $q_{1}$,  with $\phi \to +\infty$, triggered by a single trace operator (Dirichlet b.c.),
\item from $p_1$ (CFT$_{1}$ {\bf UV}) at $\phi_{*1}=0$ to $q_{1}$,  with $\phi \to +\infty$, triggered by a non-vanishing VEV (Neumann b.c.);
 \end{enumerate}
    \item  $\frac{1}{2}< a^2<1$
    \begin{enumerate}
    \item  from $p_1$ (CFT$_{1}$ {\bf UV})  at $\phi_{*1}=0$ to $p_4$, with $\phi \to +\infty$, triggered by a source (Dirichlet b.c.)
    \item from $p_1$ (CFT$_{1}$ {\bf UV})  at $\phi_{*1}=0$ to $p_4$, with $\phi \to +\infty$, triggered by a VEV of the operator (Neumann b.c.)
    
            \item  from  $p_1$ (CFT$_{1}$ {\bf UV})  to $p_2$  (CFT$_{2}$ {\bf IR}), driven by a source (Dirichet b.c.)
            \item  from  $p_1$ (CFT$_{1}$ {\bf UV})  to $p_2$  (CFT$_{2}$ {\bf IR}), driven by a non-zero VEV (Neumann b.c., Mixed b.c)
        \item from $p_1$ (CFT$_{1}$ {\bf UV})  to $p_3$   (CFT$_{3}$ {\bf IR}), driven by a source (Dirichlet b.c.)
        \item  from $p_1$ (CFT$_{1}$ {\bf UV})  to $p_3$   (CFT$_{3}$ {\bf IR}), driven by a non-zero VEV (Neumann b.c., Mixed b.c)
    \end{enumerate}
\end{enumerate}


\end{document}